\documentclass[nofootinbib]{revtex4}
\usepackage{graphicx}% Include figure files
\usepackage{dcolumn}% Align table columns on decimal point
\usepackage{amsmath,amssymb,epsfig}
\usepackage{paralist}
\usepackage{comment} 
\usepackage{graphicx}
\usepackage{multirow}
\usepackage{color,soul}
\usepackage{suffix}
\usepackage{mathtools}

\usepackage[breaklinks=True,colorlinks=True, linkcolor=magenta,citecolor=blue,debug=True]{hyperref}

\usepackage{wasysym}
\renewcommand{\vec}[1]{\boldsymbol{\mathrm{#1}}}

\def\beqn{\begin{eqnarray*}}
\def\eeqn{\end{eqnarray*}}

\newcommand{\be}{\begin{equation}}
\newcommand{\ee}{\end{equation}}
\newcommand{\ba}{\begin{eqnarray}}
\newcommand{\ea}{\end{eqnarray}}

\begin{document}

\title{Relativistic Time Transformations Between \\
the Solar System Barycenter, Earth, and Moon}

\author{Slava G. Turyshev, James G. Williams, Dale H. Boggs, and Ryan S. Park
}   

\affiliation{ 
Jet Propulsion Laboratory, California Institute of Technology,\\
4800 Oak Grove Drive, Pasadena, CA 91109-0899, USA
}%

\date{\today}% It is always \today, today,

\begin{abstract}  

Relativistic corrections are essential for time transformations between geocentric, solar system barycentric, and luni-centric reference systems to account for differences in gravitational potential and relative motion. As the primary reference for Earth-based systems, Terrestrial Time ({\tt TT}) provides the foundation for precise synchronization across spatial and temporal frameworks. To ensure consistency with {\tt TT}, Barycentric Dynamical Time ({\tt TDB}) must exhibit no average rate difference from {\tt TT}. Although the International Astronomical Union (IAU) has established resolutions for transformations between {\tt TT} and {\tt TDB}, extending these frameworks to define a lunar surface time scale ({\tt TL}) is essential for advancing lunar exploration. This paper derives the $({\tt TL}-{\tt TT})$ transformation, quantifying a secular drift of 56.0256 $\mu$s/day and periodic terms, with the largest amplitude of $\sim0.470~\mu$s at the mean anomalistic period. Additionally, the {\tt TT}-compatible spatial scale and Lorentz contraction of Moon-centered positional coordinates are computed, achieving sub-nanosecond timing precision. These transformations, implemented in JPL ephemeris generation software, provide a robust framework for high-fidelity relativistic models of lunar timekeeping, enabling further refinements and supporting navigation, communication, and scientific operations in cis-lunar space. 

\end{abstract}

\maketitle

\section{Introduction}

As our lunar presence expands, synchronizing an extensive network of assets on the Moon and in cislunar space with Earth-based systems will become increasingly challenging. To address this, it is essential to establish a common system time for all lunar assets. This system would account for relativistic effects due to differing gravitational and motion conditions, ensuring precise and efficient operations across cislunar space.
The increasing complexity of lunar exploration missions, including autonomous systems and long-duration crewed activities, requires precise synchronization between Earth-based and lunar systems. Relativistic time transformations provide the foundation for achieving this precision by addressing discrepancies caused by gravitational potential differences and relative motion.

The adoption of a dedicated lunar time and a Luni-centric Coordinate Reference System ({\tt LCRS}) is fundamental to the success of lunar exploration. These elements will ensure precise positioning, navigation, and timing (PNT), thus facilitating scientific discovery, efficient resource utilization, and the expansion of human presence on the Moon.

Although relativistic time transformations between the Solar System's Barycentric  ({\tt SSB}) coordinate reference system and the Geocentric Coordinate Reference System ({\tt GCRS}) are familiar (Chapter 10 of \cite{Petit-Luzum:2010}), an analogous transformation for the {\tt LCRS} had not been established. In particular, the constants that describe the behavior of various time scales at the Moon as time progresses are needed. This paper presents a transformation and the associated constants. 

This paper is organized as follows: Section~\ref{sec:time-E} presents the time and position transformations considering Sun, Earth, Moon, and planets other than Earth. In Section~\ref{sec:time-M} we develop similar transformation for the Moon.  Section~\ref{sec:imp} is about implementation. We explore the relationships between the lunar and terrestrial time scales, gaining insights into the structure of the relevant time transformations. We also assess the magnitude of the drift rate between the lunar and terrestrial clocks.  In Section~\ref{sec:conlude} we discuss the results obtaibed.   Appendix~\ref{sec:planet-GR} provides details on the influence of the planets.  In Appendix~\ref{sec:appA}, we present the pertinent IAU Resolutions that address the relevant relativistic modeling. In Appendix~\ref{sec:appC}, we discuss the structure of the terms relevant to the time transformation between the {\tt GCRS} and {\tt LCRS}.

\section{Time and Position Transformations for the Earth system}
\label{sec:time-E}

There is a history of papers about relativistic time transformations for the Earth \cite{Moyer:1981-part1,Moyer:1981-part2,Fairhead-Bretagnon:1990,Fukushima:1995,Irwin-Fukushima:1999,Fukushima:2010, Standish:1998,Standish_Williams_2008,Brumberg-Groten:2001}. These have resulted in and been related to a series of International Astronomical Union (IAU) Resolutions \cite{Kaplan:2005,Luzum-etal:2011}. There is a variety of time scales related to different reference systems and locations (see the IAU Resolutions in Appendix~\ref{sec:appA} and Chapter 10 of the International Earth Rotation and Reference Systems Service (IERS) Conventions \cite{Petit-Luzum:2010}). For the Earth, we want to get from International Atomic Time ({\tt TAI}) or Terrestrial Time ({\tt TT}) to Barycentric Dynamical Time ({\tt TDB}). Coordinate times  {\tt TAI}  and {\tt TT} are appropriate for the Earth’s surface on the geoid, with {\tt TT} being close to proper time on or near Earth’s surface.\footnote{While coordinate time transformations are defined between reference systems rather than physical locations, our focus is on practical applications for users on the Earth’s or Moon’s surface, which we emphasize by directly referencing a surface-based user.} Time {\tt TDB} corresponds to the {\tt SSB} frame. (See Table~\ref{tab:abbr} for a list of  abbreviations used in this paper.) After long-term averaging, the {\tt TDB} runs at the same average rate as {\tt TT} or {\tt TAI}, but there are differences that can be represented with a series of periodic terms including a large annual term with an amplitude of 1.55 milli-second (ms). 

In this section, we review the time transformation models specifically developed for the Earth system. This review is essential, as our method for introducing the {\tt LCRS} in Sec.~\ref{sec:time-M} will closely parallel the approach used for the {\tt GCRS}. 

\subsection{BCRS and GCRS, as defined by the IAU}
\label{sec:B-GCRS}

\begin{table*}[t!]
\vskip-15pt
\caption{List of abbreviations used in the paper.
\label{tab:abbr}}
\begin{tabular}{|c| l |}\hline
{\tt BCRS} & Barycentric Celestial Reference System\\
{\tt SSB} & Solar System Barycentric\\
{\tt TCB}  & Barycentric Coordinate Time \\
{\tt TDB} & Barycentric Dynamical Time \\\hline
{\tt GCRS}  & Geocentric Celestial Reference System\\
 {\tt TCG} & Geocentric Coordinate Time \\
  {\tt TT} & Terrestrial Time\\
  \hline
 {\tt LCRS} &  Luni-centric Celestial Reference System \\
 {\tt TCL}&  Luni-centric Coordinate Time\\
 {\tt TL} & Lunar Time\\
 \hline
 IAU & International Astronomical Union \\
{\tt TAI} & International Atomic Time\\
\hline
\end{tabular}
\end{table*}

The IAU 2000 Resolutions  \cite{Kaplan:2005} have presented relativistic formulation for the {\tt BCRS} and the {\tt GCRS} that ensures an uncertainty of no greater than $5 \times 10^{-18}$ in time rate and 0.2 ps in time for locations beyond a few solar radii from the Sun \cite{Soffel-etal:2003}. However, for the near-term lunar applications we will not need such an exquisite precision.  In fact, modeling uncertainties of $\sim1 \times 10^{-16}=8.64$~picosecond/day (ps/d) in time rate  and $\sim10$~ps for time will be sufficient. Therefore, below, we will refer to these accuracy assumptions when we develop the relevant expressions. 

For practical purposes, one needs a chain of time transformations from {\tt TT} to Geocentric Coordinate Time ({\tt {\tt TCG}}) in the Geocentric Celestial Reference System ({\tt GCRS}),  to  Barycentric Coordinate Time ({\tt TCB}) in the Barycentric Celestial Reference System ({\tt BCRS}), and to {\tt TDB} in the {\tt SSB} frame. One also needs inverse time transformations from {\tt TDB} to {\tt TT}. 

According to the IAU 2000 Resolutions \cite{Soffel-etal:2003,Kaplan:2005}, the metric tensor of the {\tt BCRS} with coordinates ($t$ = $\tt TCB$, $\vec x$) may be given in the following form accurate to ${\cal O}(c^{-4})$, sufficient for our purposes:
{}
\begin{equation}
ds^2_{\tt BCRS} = \Big\{1 - 2c^{-2}\sum_{\tt B} U_{\tt B}(t, \vec x) +{\cal O}(c^{-4})\Big\}c^2dt^2 - \Big\{1 + 2c^{-2}\sum_{\tt B} U_{\tt B}(t,\vec x)+{\cal O}(c^{-4})\Big\} d\vec x^2,
\label{eq:metric-CB}
\end{equation}
where $U_{\tt B}$ is the Newtonian potential of a body {\tt B} and summation is over all the solar system bodies. The omitted ${\cal O}(c^{-4})$ terms in this expression, when evaluated at the Earth,  contribute up to  $\sim 9.74\times 10^{-17}=8.42$~picoseconds/day (ps/d), too small to consider for the anticipated timing accuracy.

Similarly, the {\tt GCRS} with coordinates ($T$ = $\tt TCG$, $\vec X$) is described by a metric that 
{}
\begin{equation}
ds^2_{\tt GCRS} = \Big\{1 - 2c^{-2}\Big(U_{\tt E}(T, \vec  X) + U_{\rm tid}(T, \vec  X)\Big)   +{\cal O}(c^{-4})\Big\}c^2dT^2 - \Big\{1 + 2c^{-2}\Big(U_{\tt E}(T, \vec  X) + U_{\rm tid}(T, \vec  X)\Big)+{\cal O}(c^{-4})\Big\} d\vec X^2,
\label{eq:metric-C}
\end{equation}
where  $U_{\tt E}(T, {\bf X})$ is the Newtonian gravitational potential  of the isolated Earth and $U_{\rm tid}(T, {\bf X})$ is  the tidal potential produced by all other solar system bodies (excluding Earth). The omitted terms are due to $U_{\tt E}^2$ that may contribute up to  $\sim 8.43\times 10^{-19}$, negligible for our purposes.

Finally, the IAU 2000 Resolutions (see \cite{Soffel-etal:2003,Petit-Luzum:2010}) also provided the time and position transformations between the {\tt GCRS} ($T$ = $\tt TCG$, $\vec X$) and the {\tt BCRS} ($t$ = $\tt TCB$, $\vec x$) that, to sufficient accuracy, may be given as follows:
{}
\begin{eqnarray}
T &=& t-c^{-2}\Big\{\int^{t}_{t_{0}}\Big( {\textstyle\frac{1}{2}}v_{\tt E}^2+\sum_{{\tt B}\not= {\tt E}} U_{\tt B}  \Big)dt + (\vec v_{\tt E} \cdot \vec r_{\tt E}) \Big\} +
 {\cal O}\big(1.10\times 10^{-16}\big)(t-t_0),
 \label{eq:coord-tr-T1-rec}\\[4pt]
\vec X &=&\Big(1+  c^{-2}\sum_{{\tt B}\not= {\tt E}} U_{\tt B}\Big) \vec r_{\tt E}+ c^{-2}{\textstyle\frac{1}{2}}( \vec v_{\tt E} \cdot\vec r_{\tt E})\vec v_{\tt E}  + {\cal O}\big(1.34\times 10^{-6}~{\rm m}\big),
\label{eq:coord-tr-Xrec}
\end{eqnarray}
where $\vec r_{\tt E} \equiv \vec x - \vec x_{\tt E}(t)$ with $\vec x_{\tt E}$  and $\vec v_{\tt E}=d\vec x_{\tt E}/dt$ being the Earth's position and velocity vectors  in the {\tt BCRS}. Note that in their most accurate form (see \cite{Soffel-etal:2003}), the time transformations  (\ref{eq:coord-tr-T1-rec}) also include terms $\propto{\cal O}(c^{-4})$, while  the spatial transformation (\ref{eq:coord-tr-Xrec})  contains  acceleration-dependent terms. Taking this into account, the  uncertainty in (\ref{eq:coord-tr-T1-rec}) comes from the omitted ${\cal O}(c^{-4})$ terms behaving as $\sim c^{-4}\big\{-\frac{1}{8}v_{\tt E}^2-\frac{3}{2}v^2_{\tt E}G M_\odot/r_{\tt E}+\frac{1}{2}(G M_\odot/r_{\tt E})^2\big\}\lesssim -1.10\times 10^{-16}=-9.50$~ps/d.  Similarly, the  uncertainty in  (\ref{eq:coord-tr-Xrec}) is set by the omitted acceleration-depended terms that on the Earth's surface may contribute up to $c^{-2}\big((\vec a_{\tt E}\cdot \vec r_{\tt E})\vec r_{\tt E}- \frac{1}{2}r^2_{\tt E}\vec a_{\tt E}\big)\simeq 1.34\times 10^{-6}$~m. Even at the lunar distance, this term contributes only $\sim4.87 \times 10^{-3} \, \text{m}$, which is negligible for our purposes. Eqs.~(\ref{eq:coord-tr-T1-rec})--(\ref{eq:coord-tr-Xrec}) will be used to develop {\tt TCG} to {\tt TCB}  time scales.   

\subsection{Relativistic time scales}
\label{sec:GCRS}

We first consider  the relationship between $\tt TT$  and $\tt TCG$. Time $\tt TT$ was defined by IAU Resolution A4 (1991) \cite{Guinot:1992} as: a time scale differing from $\tt TCG$ by a constant rate, with the unit of measurement of $\tt TT$ chosen so that it matches the SI second on the geoid. According to the transformation between proper and coordinate time from (\ref{eq:metric-C}) that yields $d{\tt TT}=\big(1 - c^{-2}(U_{\tt E} + U_{\rm tid}+{\textstyle\frac{1}{2}} \dot X^2)+{\cal O}(c^{-4})\big)d {\tt TCG} $, this constant rate is expressed as 
{}
\begin{equation}
\Big<\frac{d{\tt TT}}{ d {\tt TCG}}\Big> = 1 - \frac{1}{c^2}\big<W_{\tt gE} \big>
= 1 - L_{\tt G}, 
\label{eq:LG}
\end{equation}
where $W_{\tt gE}$ is the combined gravitational and rotational potential on the geoid, determined as $W_{\tt gE}=(62636856.0\pm0.5)~{\rm m}^2{\rm s}^{-2}$ \cite{Groten:2004}. The IAU value for $L_{\tt G}$ is $6.969\,290\,134 \times 10^{-10}\approx 60.2147$ microseconds/day $(\mu{\rm s/d})$,  a defining constant as set by IAU 2000 Resolution B1.9, Table 1.1 in \cite{Petit-Luzum:2010}. It can be calculated based on the potential at the geoid, specifically at the equator, and the kinetic energy due to Earth's rotation:
{}
\begin{equation}
L_{\tt G} \equiv \frac{1}{c^2}\Big<W_{\tt g E}\Big>=\frac{1}{c^2}\Big<U_{\tt E}+{\textstyle\frac{1}{2}} [\vec \omega_{\tt E}\times \vec R_{\tt E}]^2 \Big>\simeq \frac{1}{c^2}\Big\{\frac{GM_{\tt E}}{R_{\tt E}} ( 1 + {\textstyle\frac{1}{2}}J_2 ) +  {\textstyle\frac{1}{2}} R_{\tt E}^2 \omega_{\tt E}^2 \Big\}+{\cal O}(3.29 \times 10^{-15}),
\label{eq:(5)}
\end{equation}
where $GM_{\tt E} = 398600.44\, {\rm km}^3/{\rm s}^2$ and the equatorial radius of the Earth is $R_{\tt E} = 6378.1366$ km. The un-normalized zonal harmonic coefficient $J_2$ is $1.0826359\times 10^{-3}$, as listed in Table 1.1 of \cite{Petit-Luzum:2010}, and the Earth's rotation rate $\omega_{\tt E}$ is $1.0027379$~rev/day or $\omega_{\tt E}=7.29212\times 10^{-5}~{\rm s}^{-1}$. The speed of light, $c$, is defined as $299792.458$ km/s according to \cite{Mohr-etal:2016}.  Based on these numbers, our computed value for $L_{\tt G}$ is $6.969\,283\,836 \times 10^{-10}\approx 60.2146~(\mu{\rm s/d})$ or $ - 0.6\times 10^{-15}$ (about $ - 0.0001\%$) different from the IAU value (see Table~\ref{tab:1}). Note that the IAU value was computed using the $W_{\tt gE}$  from \cite{Groten:2004}, while in our case,  the solar system ephemerides \cite{Park-etal:2021} were used. Also, the error bound in (\ref{eq:(5)}) is set by the omitted term with the tesseral harmonics $C_{22}=1.5745\times 10^{-6}$ of the Earth's gravity field \cite{Montenbruck-Gill:2012}. See Ref.~\cite{Turyshev-Toth:2023-grav-phase} for a definition of $W_{\tt gE}/c^2$ in (\ref{eq:(5)}) accurate to ${\cal O}(5.83\times 10^{-17})$, which includes gravitational harmonics $J_\ell, C_{\ell k}$ and $S_{\ell k}$  up to $\ell=8$ order.

The tidal terms from (\ref{eq:metric-C})  were evaluated  to contribute only at the levels of approximately $4.90\times 10^{-17}$ for the Moon and $1.79\times 10^{-17}$ for the Sun \cite{Turyshev-Toth:2023-grav-phase}, which is beyond the stated accuracy limit. Thus, they were omitted below.

Note that, if based on geoid (\ref{eq:LG}), the definition of $\tt TT$ is accurate only to about $10^{-17}$ \cite{Soffel-etal:2003,Kaplan:2005}. The intricacy and temporal variability associated with the definition and realization of the geoid contribute to uncertainties in defining and realizing $\tt TT$. Recognizing these challenges, the constant $L_{\tt G}$ was turned into a defining constant with its value fixed to $6.969\,290\,134 \times 10^{-10}$ (2000 IAU Resolution B1.9) \cite{Soffel-etal:2003,Kaplan:2005}, see Table~\ref{tab:1}.

The conversion from {\tt TT} to Geocentric Coordinate Time ({\tt TCG}), on average, involves a rate change  (\ref{eq:LG})
{}
\begin{equation}
\frac{	d {\tt {\tt TCG}}}{d {\tt TT}} = \frac{1}{1  -  L_{\tt G}} = 1 + \frac{L_{\tt G}}{1 -  L_{\tt G}},
\label{eq:(4)}
\end{equation}
which may be used to introduce the following  relationship between  {\tt TCG} and {\tt TT}, starting at time ${\tt T}_0$:
{}
\begin{equation}
 {\tt TCG}-{\tt TT}  = \frac{L_{\tt G}}{1 -  L_{\tt G}}({\tt TT}-{\tt T}_0).
\label{eq:(4)in}
\end{equation}

According to the IAU Resolution B1.9 (2000), $L_{\tt G}$, previously dependent on the geoid model as per IAU (1991) resolutions, is no longer considered variable. Furthermore, the scaling of spatial coordinates and mass factors is designed to maintain the invariance of the speed of light and the equations of motion in the {\tt GCRS} \cite{Klioner:2008}, applicable to the Moon's or Earth’s artificial satellites, during the transformation from {\tt TCG} to {\tt TT}. This transformation, which includes the scaling of temporal and spatial coordinates and mass factors, ensures the invariance of the metric (up to a constant factor)
{}
\begin{equation}
(ds^2)_{\tt TT} = (1- L_{\tt G})^2ds^2_{\tt TCG} ,
\label{eq:interv}
\end{equation}
where $(ds^2)_{\tt TT}$ maintains the same form in terms of {\tt TT}, $\vec X_{\tt TT}$, and $(G M )_{\tt TT}$ as  (\ref{eq:metric-C}) does in terms of $T$, $\vec X$, and $(GM)_{\tt TCG}$. 

As a result, instead of coordinate time ${\tt T} = {\tt TCG}$, spatial coordinates $\vec X$ and mass factors $(G M)_{\tt TCG}$ related to {\tt GCRS}, the following scaling of these quantiles is used \cite{Brumberg-Groten:2001}
{}
\begin{equation}
{\tt TT} ={\tt TCG}- L_{\tt G}({\tt TCG}-{\tt T}_0), \qquad
\vec X_{\tt TT} = (1-L_{\tt G})\vec X_{\tt TCG}, \qquad
(GM)_{\tt TT} = (1-L_{\tt G})(GM)_{\tt TCG}.
\label{eq:TCGT}
\end{equation}

We note that, using (\ref{eq:metric-C}), the proper time \( \tau_{\tt C} \) measured by a clock located at the {\tt GCRS} coordinate position \( \vec{X}_{\tt C}(T) \) and moving with the coordinate velocity \( \vec{v}_{\tt C} = d{\vec X}_{\tt C}/dT \), can be expressed as:
{}
\begin{equation}
\frac{d\tau_{\tt C}}{d{\tt TCG}}= 1-\frac{1}{c^2}\Big\{ {\textstyle\frac{1}{2}}\vec{v}_{\tt C}^2+U_{\tt E}(T, \vec  X_{\tt C}) + U_{\rm tid}(T, \vec  X_{\tt C})\Big\}_{\tt TCG}+{\cal O}(c^{-4}).
\label{eq:proper-t-C-comp}
\end{equation}

Using (\ref{eq:proper-t-C-comp}) together with (\ref{eq:(4)}), we may express $\tau_{\tt C}$ as a function of {\tt TT} as below
{}
\begin{equation}
\frac{d\tau_{\tt C}}{d{\tt TT}}=\frac{d\tau_{\tt C}}{d{\tt TCG}}\frac{d{\tt TCG}}{d{\tt TT}}\simeq 1+L_{\tt G}-\frac{1}{c^2}\Big\{{\textstyle\frac{1}{2}}{\vec v}^2_{\tt C}
+ U_{\tt E}({\vec X}_{\tt C})+ U_{\rm tid}(\vec  X_{\tt C})\Big\}_{\tt TT}+{\cal O}(c^{-4}).
\label{eq:synch}
\end{equation}

Clearly, if the target clock synchronization is of the order of  $\sim 10^{-15}$, the definition (\ref{eq:synch}) is rather clean with just a few mutipole terms \cite{Turyshev-Toth:2023-grav-phase}. This expression quickly becomes rather messy if a more precise synchronization is desired.

Another constant, $L_{\tt C}$, removes the average rate between ${\tt TCG}$ and ${\tt TCB}$ given by (\ref{eq:coord-tr-T1-rec}). It is determined as the long time average of the rate computed from  transformation (\ref{eq:coord-tr-T1-rec}) given as below
{}
\begin{equation}
{\tt TCG} -{\tt TCB} = -\frac{1}{c^2} \Big\{\int \Big( {\textstyle\frac{1}{2}}v_{\tt E}^2 + \sum_{{\tt B}\not={\tt E}}U_{\tt B} \Big) d{\tt TCB}  + (\vec v_{\tt E} \cdot \vec r_{\tt E})\Big\}_{\tt TCB} +{\cal O}(c^{-4}),
\label{eq:(12)}
\end{equation}
or the inverse to it which given as 
{}
\begin{equation}
{\tt TCB} - {\tt TCG} = \frac{1}{c^2}\Big\{ \int \Big( {\textstyle\frac{1}{2}}v_{\tt E}^2 + \sum_{{\tt B}\not={\tt E}}U_{\tt B} \Big) d{\tt TCG} +(\vec v_{\tt E} \cdot \vec X)\Big\}_{\tt TCG} +{\cal O}(c^{-4}),
\label{eq:(12inv)}
\end{equation}
where the subscripts $\{...\}_{\tt TCB}$ and $\{...\}_{\tt TCG}$ are used to identify ${\tt TCB}$- or ${\tt TCG}$-compatible quantities. Similar notations will be used below to highlight the use of other relevant time scales.

\begin{table*}[t!]
\vskip-15pt
\caption{Constant terms for Earth $d{\tt TDB}/d{\tt TT}$ and Moon $d{\tt TDB}/d{\tt TL}$. Quantities $U_{\tt ES}$  and $U_{\tt EM}$ are the gravitational potentials of the Sun and Moon, evaluated at the origins of the {\tt GCRS}, while $U_{\tt MS}$  and $U_{\tt ME}$ are similar quantities due to Sun and Earth  at the origin of the {\tt LCRS}.
Quantity $U_{\tt BP}$ is the gravitational contribution at Earth-Moon barycenter from the planets at J2000 is given in Table~\ref{tab:2} of Appendix \ref{sec:planet-GR}. The constant $L_{\tt L}$ (analogous to $L_{\tt G}$ for Earth) accounts for the average rate of time transformation between the Moon's center and the vicinity of its surface, given by (\ref{eq:(26)}), (\ref{eq:(29)}). The constant $L_{\tt H}$ (analogous to $L_{\tt C}$) represents the long-time average of the Moon's total orbital energy in its motion around the {\tt BCRS}, given by (\ref{eq:coord-tr-QQM}). The constant $L_{\tt M}$ (analogous to $L_{\tt B}$) compensates for the average rate of time transformation between the {\tt TCB} and Lunar Time ({\tt TL}) at the surface of the Moon (analogous to {\tt TT}), given by (\ref{eq:LM}), (\ref{eq:(nonN4)}). Finally, the constant \( L_{\tt EM}= L_{\tt H}-L_{\tt C} \) represents the long-time average of the Moon's total orbital energy in its motion around the Earth, as observed from the {\tt GCRS}, and is introduced in  (\ref{eq:expRR1+}).  
\label{tab:1}}
\begin{tabular}{|c|l|}\hline
Parameter &~~~~Value, $10^{-8}$ \\\hline
$U_{\tt ES}/c^2$ 	&~ 0.9870628  \\
$U_{\tt EM}/c^2$ 	&~  0.0000142 \\
$U_{\tt BP}/c^2$	&~  0.0002197 \\
${\textstyle\frac{1}{2}}v_{\tt E}^2/c^2$ 
&~  0.4935299 \\
$L_{\tt B}$, IAU 	&~  1.550519768 \\
$L_{\tt B}$, this paper	&~  1.55051969313 \\
$L_{\tt G}$, IAU 	&~  0.06969290134  \\
$L_{\tt G}$, this paper 	&~  0.06969283836 \\
$L_{\tt C}$, IAU	&~ 1.48082686741 \\
$L_{\tt C},$ this paper &~ 1.48082685455 \\
\hline
$U_{\tt MS}/c^2$ 	&~  0.9870647 \\
$U_{\tt ME}/c^2$ 	&~  0.0011538  \\
$U_{\tt BP}/c^2$ 	&~  0.0002197 \\
${\textstyle\frac{1}{2}}v_{\tt M}^2/c^2$  &~  0.4941002 \\
$L_{\tt L}$ 	&~  0.003139054 \\
$L_{\tt H}$ 	&~  1.48253624 \\
$L_{\tt M}$ 	&~  1.48567529 \\
$L_{\tt EM}= L_{\tt H}-L_{\tt C}$ & ~~0.001709385 \\
\hline
\end{tabular}
\end{table*}

Although the integrals in (\ref{eq:(12)}), (\ref{eq:(12inv)})  may be calculated by a numerical integration (see details in \cite{Fukushima:1995,Irwin-Fukushima:1999}), there are analytic formulations available (e.g., \cite{Fairhead-Bretagnon:1990,Harada-Fukushima:2003}). For that, expression for the the total Earth's energy of its orbital motion may be given as below:
{}
\begin{eqnarray}
\frac{1}{c^2}\Big({\textstyle\frac{1}{2}}v_{\tt E}^2 +  \sum_{{\tt B}\not={\tt E}}U_{\tt B} \Big)=L_{\tt C}+ \dot P(t)+{\cal O}(c^{-4}), \qquad 
\frac{1}{c^2}\Big<{\textstyle\frac{1}{2}}v_{\tt E}^2 +  \sum_{{\tt B}\not={\tt E}}U_{\tt B} \Big>=L_{\tt C},
 \label{eq:coord-tr-QQ}
\end{eqnarray}
where $\left<...\right>$ denotes the long time average. 
The constant \( L_{\tt C}\) is derived from long-term averaging of Earth's total orbital energy, as expressed in (\ref{eq:coord-tr-QQ}), yeilding \( L_{\tt C} = 1.480\,826\,854\,55 \times 10^{-8} \approx 1.279\,434\,4~\text{ms/d} \) (milliseconds per day). The term \( P(t) \) represents a series of periodic components, as detailed in Refs.~\cite{Fairhead-Bretagnon:1990,Irwin-Fukushima:1999}. Notably, the largest term in the \( P \)-series has an amplitude of \( A_1 = 1656.674564 \times 10^{-6}~\text{s} \) and a frequency of \( \omega_1 = 6283.079543033~\text{rad}/(10^3~\text{yr}) \simeq 1.9923515 \times 10^{-7}~\text{s}^{-1} \), as reported in \cite{Fairhead-Bretagnon:1990,Petit-Luzum:2010,Harada-Fukushima:2003}.

The IAU 2000 Resolution B1.5 (Appendix~\ref{sec:appA}) uses a slightly different value of \( L_{\tt C} = 1.480\,826\,867\,41 \times 10^{-8} \pm 2 \times 10^{-17} \approx 1.279\,434\,4~\text{ms/d} \pm 1.7~\text{ps/d} \), introducing it as a fundamental constant\footnote{See the IAU 2009 System of Astronomical Constants: \url{https://iau-a3.gitlab.io/NSFA/IAU2009_consts.html\#g2_16}} (refer to Table 1.1 in \cite{Petit-Luzum:2010}). The discrepancy between our value and the IAU-adopted value arises from the use of different ephemerides.

As a result, equation (\ref{eq:(12)}) may be used to determine mean rate between ${\tt TCG}$ and ${\tt TCB}$:
{}
\begin{equation}
\Big<\frac{d{\tt TCG}}{d{\tt TCB}}\Big> =1-L_{\tt C}.
\label{eq:constTCGTCB}
\end{equation}

Similar to (\ref{eq:LG}), we can formally relate {\tt TT} and {\tt TCB}. Using (\ref{eq:metric-CB}),  one may derive the transformation between time on or near the surface of the Earth and the relevant coordinate time, as was done in (\ref{eq:LG}). Then, the constant resulting from this, $L_{\tt B}$, would be a long-term average of the total potential and kinetic energy at the clock's location, yielding the following definition for $L_{\tt B}$ as the constant rate between {\tt TT} and {\tt TCB}:
{}
\begin{equation}
\Big<\frac{d{\tt TT}}{ d {\tt TCB}}\Big> = 1 - L_{\tt B}.
\label{eq:LB}
\end{equation}

However, implementing such a definition in practice is challenging due to the time-varying contributions of planetary motions (see discussion in Sec.~\ref{sec:planet-GR}). To address this, the IAU 2000 Resolutions B1.5 and B1.9 (see Appendix~\ref{sec:appA}) establish a framework to relate the constants \( L_{\tt B} \), \( L_{\tt G} \), and \( L_{\tt C} \), enabling the inference of \( L_{\tt B} \). This approach complemented by IAU 2006 Resolution B3 for {\tt TDB} and $L_{\tt B}$ \cite{Luzum-etal:2011} defines the relationship between \( {\tt TDB} \) and \( {\tt TCB} \) using the constant \( L_{\tt B} \) while ensuring there is no rate difference between \( {\tt TDB} \) and \( {\tt TT} \):
{}
\begin{equation}
\Big<\frac{d{\tt TDB}}{d{\tt TCB}}\Big>= 1-L_{\tt B} \qquad {\rm and}\qquad 
\frac{d{\tt TDB}}{d{\tt TT}}=1.
\label{eq:const=}
\end{equation}
Using these expressions together with (\ref{eq:LG}) and (\ref{eq:constTCGTCB}), we have
{}
\begin{equation}
\Big<\frac{d{\tt TDB}}{d{\tt TCB}}\Big>=\Big(\frac{d{\tt TDB}}{d{\tt TT}}\Big)\Big<\frac{d{\tt TT}}{d{\tt TCG}}\Big>\Big<\frac{d{\tt TCG}}{d{\tt TCB}}\Big> \qquad
\Rightarrow \qquad 1-L_{\tt B}=(1-L_{\tt G})(1-L_{\tt C}),
\label{eq:constLBLCLG}
\end{equation}
where $L_{\tt B}$ is determined as $L_{\tt B} = L_{\tt G}+L_{\tt C}-L_{\tt G}L_{\tt C}=
1.550\,519\,768 \times 10^{ -8}\pm 
2\times 10^{ - 17}
\approx 1.339\,65~{\rm ms/d}\pm 1.7$~ps/d,  an IAU defining constant \cite{Kaplan:2005,Petit-Luzum:2010,Luzum-etal:2011}. Table~\ref{tab:1} summarizes various constants in the rate equations for Earth and Moon. 

To estimate the value of $L_{\tt B}$, while  keeping (\ref{eq:LB}), we may use
{}
\begin{align}
L_{\tt B} = \frac{1}{c^2}\Big\{ GM_{\tt S} \Big< \frac{1}{r_{\tt ES}}\Big> + GM_{\tt M} \Big< \frac{1}{r_{\tt EM}}\Big> + \big<U_{\tt EP}\big> + \big<{\textstyle\frac{1}{2}}v_{\tt E}^2\big>+ \frac{GM_{\tt E}}{R_{\tt E}} ( 1 + {\textstyle\frac{1}{2}}J_{2\tt E}) + {\textstyle\frac{1}{2}}R_{\tt E}^2 \omega_{\tt E}^2  \Big\} +{\cal O}(3.29 \times 10^{-15}),
\label{eq:(nonN3)}
\end{align}
where $\left<...\right>$ denotes a time average, $\left<U_{\tt EP}\right>$ is the sum of the potentials of the planets other than Earth, presented in Appendix~\ref{sec:planet-GR}.  This results in the value $L_{\tt B}= 1.550\,519\,693\,13 \times 10^{ - 8}\approx 1.339\,65$~ms/d. Even then, the difference between the IAU value and the value computed here is $3\times 10^{-15}$, larger than the uncertainty implied by the number of digits in  \cite{Petit-Luzum:2010}. This correction, while minor, is critical for applications requiring sub-nanosecond timing accuracy, such as relativistic transformations in ephemeris generation and synchronization frameworks.  The equatorial radius of the Earth was set at $R_{\tt E} = 6378.1366$~km  \cite{Petit-Luzum:2010}. The GM values for the Sun, Earth, Moon, and planets were taken from DE440 \cite{Park-etal:2021}.  The values in Table~\ref{tab:1} calculated in this paper are given to $10^{-15}$, which would be $0.3~\mu$s after 10 yr. These will not exactly match the values used for the IAU value of ${L_{\tt B}}$.  Similar to (\ref{eq:(5)}), the error bound in (\ref{eq:(nonN3)}) is due to the omitted term with the tesseral harmonics $C_{22}$ of the Earth's gravity field.

Recognizing the challenges introduced by temporal behavior of various quantities involved in (\ref{eq:(nonN3)}), especially when higher precision for $L_{\tt B}$ is needed, since IAU Resolution B3 (2006), $L_{\tt B}$ is a defining constant, see Table~\ref{tab:1}. Note that Table~\ref{tab:1} also contains values of $L_{\tt G}$ and $L_{\tt B}$ that were recomputed using the solar system ephemerides and a long term averaging procedure (see Appendix~\ref{sec:planet-GR}). This was done to emphasize  temporally-varying nature of these quantities.
We retained these recalculated values to ensure consistency with DE440 ephemerides, validate our methodology, and address potential discrepancies introduced by earlier constants derived from DE405, as detailed in this study.    

The scaling of spatial coordinates and mass factors is implemented to maintain the invariance of the speed of light and the equations of motion of the solar system bodies during the transformation from {\tt TCB} to {\tt TDB}. This transformation also involves the following relationship:
{}
\begin{equation}
(ds^2)_{\tt TDB} = (1- L_{\tt B})^2ds^2_{\tt TCB}, 
\label{eq:interv-B}
\end{equation}
where $(ds^2)_{\tt TDB}$ keeps the same form in terms of {\tt TDB}, $\vec x_{\tt TDB}$, and $(GM)_{\tt TDB}$ as (\ref{eq:metric-C})  does in terms of $t$, $\vec x$, and $GM$. 

As a result, {\tt TDB} is a timescale rescaled from {\tt TCB}, as defined by IAU 2006 Resolution B3 and  IAU 2009 Resolution 3 \cite{IAU2009ResB3,Petit-Luzum:2010}, given by the following set of expressions:
{}
\begin{equation}
{\tt TDB} = {\tt TCB} - L_{\tt B}({\tt TCB}-{\tt T}_0)+{\tt TDB}_0, \qquad
\vec x_{\tt TDB} = (1-L_{\tt B})\vec x_{\tt TCB}, \qquad
(GM)_{\tt TDB} = (1-L_{\tt B})(GM)_{\tt TCB},
\label{eq:TDBC}
\end{equation}
where $L_{\tt B} = 
1.550\,519\,768 \times 10^{ - 8}$, ${\tt T}_0=2443144.5003725$ Julian days, and ${\tt TDB}_0 = -65.5~\mu$s are defining constants, equivalent to those used in JPL planetary ephemerides DE405. This definition ensures that {\tt TDB} maintains the same rate as {\tt TT}  at the geocenter. The constant term ${\tt TDB}_0$ is selected to ensure reasonable consistency with the widely used $(\tt TDB - TT)$ formula from \cite{Fairhead-Bretagnon:1990}. Note that the inclusion of ${\tt TDB}_0$ implies that {\tt TDB} is not synchronized with {\tt TT}, {\tt TCG}, and {\tt TCB} as of 1 January 1977 at 0\,h 32.184\,s {\tt TAI} at the geocenter. Additionally, we note that the IAU definition of {\tt TDB} is based on the JPL ephemeris DE405. For more details, see \url{http://www.iaufs.org/res.html}.

As shown in Table~\ref{tab:1}, there is a difference of \( \sim 7.49 \times 10^{-16} \) between the updated \( L_{\tt B} \), based on DE440, and the IAU value based on DE405. This discrepancy introduces a drift of $-64.7$~\text{ps/d}. The broader impact of these changes on ephemerides and timekeeping will be addressed in a separate study.

\subsection{Transformation {\tt TAI} $\rightarrow$ {\tt TDB}}
\label{sec:tai-tdb}

For practical purposes, {\tt TT} is expressed using International Atomic Time ({\tt TAI}) with a constant offset:
{}
\begin{equation}
{\tt TT} = {\tt TAI} + 32.184 \,{\rm s}. 	
\label{eq:(3)}
\end{equation}

Consider the three transformations for {\tt TT} $\rightarrow$ {\tt TDB} that includes the following chain {\tt TT} $\rightarrow$ {\tt TCG}, {\tt TCG} $\rightarrow$ {\tt TCB}, and {\tt TCB} $\rightarrow$ {\tt TDB}. Transformations {\tt TT} $\rightarrow$ {\tt {\tt TCG}} and {\tt TCB} $\rightarrow$ {\tt TDB} are liner in time and are given by (\ref{eq:(4)in}) and (\ref{eq:TDBC}), correspondingly. 

To establish relationships between ${\tt TCB}$ and ${\tt {\tt TCG}}$, we evaluate (\ref{eq:(12inv)}).  As we want to express this equation as a function of {\tt TT}, we need to transform its right-hand side. This can be done by using  (\ref{eq:(4)}) to express $d{\tt TCG}=d{\tt TT}/(1-L_{\tt G})$. Next, from  (\ref{eq:TCGT}), we have $\vec X_{\tt TCG}=\vec X_{\tt TT}/(1-L_{\tt G})$. With this, we  use (\ref{eq:coord-tr-QQ}) and integrate (\ref{eq:(12inv)})  over {\tt TT} from ${\tt T}_0$ to ${\tt TT}$ to present $({\tt TCB} - {\tt TCG})$ as a function of {\tt TT} as below
{}
\begin{equation}
{\tt TCB} - {\tt {\tt TCG}} = \frac{1}{1-L_{\tt G}}\Big\{L_{\tt C}\, ({\tt TT}-{\tt T}_0)+P({\tt TT})-P({\tt T}_0) +\frac{1}{c^2} (\vec v_{\tt E} \cdot \vec X_{\tt TT} )\Big\}+{\cal O}(c^{-4}),
\label{eq:(12)TT}
\end{equation}
which comprises a secular term $\propto L_{\tt C}/(1-L_{\tt G})=1.480\,826\,855 \times 10^{-8} \simeq 1.279~{\rm ms/d}$ and a series of periodic terms $P({\tt TT})$, as specified in the IERS Technical Note 36 \cite{Petit-Luzum:2010}, see also \cite{Fairhead-Bretagnon:1990,Harada-Fukushima:2003}. 

To establish relationship between  {\tt TDB} and {\tt TT}, we have the following chain of time transformations: ${\tt TDB}  -  {\tt TT}=({\tt TDB}  -  {\tt TCB})+({\tt TCB}  -  {\tt TCG})+({\tt TCG}  -  {\tt TT})$, where   $({\tt TDB}  -  {\tt TCB})$ is from (\ref{eq:TDBC}), $({\tt TCB}  -  {\tt TCG})$ is given by (\ref{eq:(12inv)}),  $({\tt TCG}  -  {\tt TT})$ is known from (\ref{eq:(4)in}), and $L_{\tt B} = L_{\tt G}+L_{\tt C}-L_{\tt G}L_{\tt C}$ from (\ref{eq:constLBLCLG}), which results in the following expression
{}
\begin{align}
{\tt TDB}  -  {\tt TT} =  {\tt TDB}_0- L_{\tt C}\,( {\tt TT}- {\tt T}_0)+\frac{1}{c^2}\Big\{\int_{\tt T_0}^{\tt TT} \Big( {\textstyle\frac{1}{2}}v_{\tt E}^2 +  \sum_{{\tt B}\not={\tt E}}U_{\tt B} \Big) 
d{\tt TT}+(\vec v_{\tt E}\cdot \vec X_{\tt TT})\Big\}  +{\cal O}(c^{-4}).
\label{eq:(nonN1)}
\end{align}

Finally, using (\ref{eq:coord-tr-QQ}), we evaluate the integral in (\ref{eq:(nonN1)}), which results in 
{}
\begin{align}
{\tt TDB}  -  {\tt TT} =  {\tt TDB}_0+P({\tt TT})-P({\tt T}_0)+
\frac{1}{c^2} (\vec v_{\tt E} \cdot \vec X_{\tt TT}) +{\cal O}(c^{-4}).
\label{eq:(nonN1T)}
\end{align}

As one can see from (\ref{eq:(nonN1T)}), introduction of {\tt TDB} in the form of (\ref{eq:TDBC}) eliminates the rate difference between {\tt TDB} and {\tt TT} at the geocenter (i.e., when $\vec X=0$), leaving only small periodic terms $\propto P({\tt TT})$. Thus, in (\ref{eq:TDBC}), the {\tt TBC} has been rescaled to become {\tt TDB} that has the same secular rate as {\tt TT}. The time {\tt TDB}  is used in the planetary and satellite ephemerides published by the JPL \cite{Park-etal:2021}. Finally, with the help of (\ref{eq:(3)}) we can express  {\tt TT} via {\tt TAI}.

\subsection{Transformation {\tt TDB}  $\rightarrow$ {\tt TAI}} 
\label{sec:tdb-tai}

Similar to (\ref{eq:(nonN1)}), we establish relationships between  {\tt TT} and {\tt TDB}, expressing the results as a function of {\tt TDB}. For that, we use the chain of time transformations: ${\tt TT}  -  {\tt TDB}=({\tt TT}  -  {\tt TCG})+({\tt TCG}  -  {\tt TCB})+({\tt TCB}  -  {\tt TDB})$, 
where relevant expressions are given by (\ref{eq:TCGT}),  (\ref{eq:(12)}), and (\ref{eq:TDBC}), correspondingly.  As a result, we have: 
{}
\begin{align}
  {\tt TT} -{\tt TDB} &=
  \frac{L_{\tt B}-L_{\tt G}}{1-L_{\tt B}}({\tt TDB}-{\tt T}_0)-
  \frac{1}{1-L_{\tt C}}\Big\{{\tt TDB}_0+\frac{1}{c^2}\int_{{\tt T_0+TDB_0}}^{\tt TDB} \Big( {\textstyle\frac{1}{2}}v_{\tt E}^2 +  \sum_{{\tt B}\not={\tt E}}U_{\tt B}\Big)d{\tt TDB}+\frac{1}{c^2}(\vec v_{\tt E}\cdot \vec {r_{\tt E}}_{\tt TDB}) \Big\} +{\cal O}(c^{-4}).
\label{eq:(nonN1-TCB)}
\end{align}
The combination of the constants $ (L_{\tt B}-L_{\tt G})/(1-L_{\tt B}) =1.480\,826\,878 \times 10^{-8}\simeq1.279~{\rm ms/d}$. This rate is removed by taking the integral with the help of (\ref{eq:coord-tr-QQ}) and  accounting for the constant $L_{\tt C}$. Indeed, evaluating the integral in (\ref{eq:(nonN1-TCB)}) together with  (\ref{eq:coord-tr-QQ}), we have the following  result that expresses {\tt TT} as a function of  {\tt TDB} (inverse to  (\ref{eq:(nonN1T)}))
{}
\begin{align}
 {\tt TT} -{\tt TDB} = -  {\tt TDB}_0-\Big\{P({\tt TDB})-P({\tt T}_0+{\tt TDB}_0)+\frac{1}{c^2} (\vec v_{\tt E} \cdot \vec {r_{\tt E}}_{\tt TDB}) \Big\} +{\cal O}(c^{-4}).
\label{eq:(nonN1TCB)}
\end{align}

Thus, there is no rate difference between  ${\tt TT}$ and  ${\tt TDB}$, only small periodic terms $\propto P({\tt TDB})$, as shown  in (\ref{eq:(nonN1T)})  and (\ref{eq:(nonN1TCB)}).
From (\ref{eq:(3)}), we know that {\tt TT} is related to {\tt TAI} by simply a reversal of that equation, yielding ${\tt TAI} = {\tt TT}  -  32.184~{\rm s}$.

\subsection{Position transformations}
\label{sec:pos-E}

The classical relativistic time transformation would include a mean rate. When we remove the mean rate, we rescale the three space coordinates as in (\ref{eq:TCGT}) and (\ref{eq:TDBC}). The other spatial effect is the Lorentz contraction along the direction of motion. Here, we  briefly address these two subjects (more relevant material is available, e.g.,  \cite{Moyer:2003,Soffel-etal:2003,Klioner:2008,Petit-Luzum:2010,Luzum-etal:2011}). 

To discuss position transformation, we consider $\vec X_{\tt TCG}$ to be a position vector of  a site on the surface of the Earth as it is observed in the {\tt GCRS} and  ${\vec r_{\tt E}}_{\tt TCB}$  being the same vector but as seen from the {\tt BCRS}. We need to express relationship between $\vec X_{\tt TCG}$ and ${\vec r_{\tt E}}_{\tt TCB}$  in terms of the {\tt TT} and {\tt TDB} timescales.  For that, we use (\ref{eq:TCGT}) to express the $\vec X_{\tt TCG}$  via ${\tt TT} $ and (\ref{eq:TDBC}) to express ${\vec r_{\tt E}}_{\tt TCB}$ via ${\vec r_{\tt E}}_{\tt TDB}$. Then, taking into account $(1-L_{\tt B})=(1-L_{\tt G})(1-L_{\tt C}),$ from (\ref{eq:constLBLCLG}), we have:
{}
\begin{eqnarray}
\vec X_{\tt TT} &=& \Big(1+ L_{\tt C} +c^{-2} \sum_{{\tt B}\not= {\tt E}} U_{\tt B} \Big){\vec r_{\tt E}}_{\tt TDB}+ c^{-2} 
\frac{1}{2}\big( \vec v_{\tt E} \cdot\vec {r_{\tt E}}_{\tt TDB}\big)\vec v_{\tt E} +
 {\cal O}\big(1.34\times 10^{-6}~{\rm m}\big),~~~~
\label{eq:coord-tr-Xrec=}
\end{eqnarray}
where $\vec r_{\tt E} \equiv \vec x - \vec x_{\tt E}(t)$  and, similar to (\ref{eq:coord-tr-Xrec}), the error here is set by the omitted acceleration-dependent terms. 

At the stated level of accuracy, developing an inverse transformation to that of (\ref{eq:coord-tr-Xrec=})  is straightforward. To achieve this, one simply needs to move the $c^{-2}$-terms  to the other side of the equation with the appropriate sign changes.

The first group of terms in (\ref{eq:coord-tr-Xrec=}) amounts to a scale change. The constant $L_{\tt C} + c^{-2}\big<\sum_{\tt B\not=E}U_{\tt B}\big>$ is ($1.48083+0.98728)\times 10^{-8} = 2.46811\times10^{-8}$ (Table~\ref{tab:1}). For the mean radius of the Earth, this contribution amounts to  $15.72$~cm. The annual variation is $\pm1.65\times 10^{-10}$, or $\pm1.05$~mm for the radius of the Earth. 
Then, to a good approximation 
{}
\begin{eqnarray}
\vec X_{\tt TT} &=& \Big(1+ 2.46811\times 10^{-8}  + 1.65\times 10^{-10} \cos l'  \Big){\vec r_{\tt E}}_{\tt TDB}+ c^{-2} \frac{1}{2}\big( \vec v_{\tt E} \cdot\vec {r_{\tt E}}_{\tt TDB}\big)\vec v_{\tt E} +
 {\cal O}\big(1.34\times 10^{-6}~{\rm m}\big),~~~~
\label{eq:coord-tr-XNN}
\end{eqnarray}
where the $\cos l'$ term accounts for the eccentricity of the annual Earth-Moon barycentric motion. 

The last term in (\ref{eq:coord-tr-XNN}) is the Lorentz contraction due to Earth's orbital motion with respect to the {\tt BCRS}. For a site on Earth's surface, the Lorentz contraction is $31.4$~mm. Along the direction of motion, the combined effect of scale change and the Lorentz contraction is $188$~mm, while perpendicular to that direction, the scale change is  $157$~mm.

Results (\ref{eq:(nonN1TCB)}), (\ref{eq:(nonN1-TCB)}) and (\ref{eq:coord-tr-Xrec=}) are relevant to most of the current practical applications. The inverse transformations to (\ref{eq:coord-tr-Xrec=}) are obtained by moving the $c^{-2}$-terms on the other side of the expression with appropriate sign changes. In Section~\ref{sec:time-M} below, we will develop similar expressions for applications in cislunar space.

\section{Time and Position Transformations for the Moon system}
\label{sec:time-M}

\subsection{Technical Basis for the LCRS}
\label{sec:tech-basis}

The Luni-centric Coordinate Reference System ({\tt LCRS}) is needed to provide a stable geospatial framework essential for accurate lunar navigation, positioning, and infrastructure planning. Although not yet formally established, the development of the {\tt LCRS} can leverage  advancements from the Gravity Recovery and Interior Laboratory (GRAIL)\footnote{See, \url{https://www.jpl.nasa.gov/missions/gravity-recovery-and-interior-laboratory-grail/}} and Lunar Reconnaissance Orbiter (LRO)\footnote{See, \url{https://science.nasa.gov/mission/lro/}} missions, as well as Lunar Laser Ranging (LLR), which collectively provide high-resolution data foundational for creating the {\tt LCRS}.

The GRAIL mission mapped the Moon’s gravitational field with unprecedented precision, reaching spherical harmonic expansions up to degree and order 1,200, with localized refinements to degree 1,500 in specific regions \cite{Konopliv-etal:2013}. This level of detail corresponds to spatial resolutions on the order of 9 km at the global scale, with some regional improvements to 7 km \cite{Zuber-etal:2013}. These models reveal gravitational anomalies, such as mass concentrations (mascons) beneath impact basins like Mare Imbrium and Mare Crisium, which cause localized gravitational highs ranging from 80 to 120 mGal. Such anomalies, if not modeled accurately, introduce trajectory perturbations, underscoring the necessity of precise gravitational data for a reliable {\tt LCRS} \cite{Konopliv-etal:2013}.

Analysis of crustal thickness from GRAIL and seismic data shows variations from 5 km in mare regions to over 70~km in the highlands, with an average crustal density of $(2,550 \pm 50)$~kg/m\(^3\) \cite{Wieczorek-etal:2013}. These structural details, along with tidal deformation data, suggest a partially molten core with a radius of 330--360 km and a density of  $\approx(5,200 \pm 200)$~kg/m\(^3\) \citep{Williams-etal:2014}. These mass distribution estimates are crucial for developing accurate gravitational models that support the {\tt LCRS}.

The lunar core structure significantly impacts the tidal Love numbers $k_2 = 0.02405 \pm 0.00018$ and $k_3 = 0.0089 \pm 0.0021$, which describe the Moon’s response to Earth’s gravitational pull and provide insights into the elasticity and structure of the lunar interior \cite{Williams-etal:2014}. The degree-2 Love number $k_2$ reflects the Moon’s relative rigidity, contrasting with Earth’s $k_2 \approx 0.3$, while the smaller degree-3 Love number $k_3$ further characterizes higher-order tidal responses. Earth’s gravitational influence induces a bulge of $\sim 10$~cm on the Moon, which, combined with librational motions, introduces periodic positional shifts that must be modeled within the {\tt LCRS} to maintain accuracy. These tidal and librational effects, which cause displacements of up to 20 meters over extended periods, necessitate ongoing adjustments within the {\tt LCRS} framework \citep{Matsumoto-etal:2010, Williams-etal:2014}.

The rotational dynamics of the Moon introduce additional complexities for the {\tt LCRS}. Earth’s gravitational influence causes a secular drift of $\sim$ \(0.02''\) per year in both longitude and latitude. The Moon’s 18.6-year precession cycle induces nutational displacements of 3--4 meters, while forced librations from orbital eccentricity produce longitudinal amplitudes around $75''$, with free librations adding shifts of about $3''$. Combined, these effects can displace reference points by 10--20 meters over time, requiring regular adjustments within the {\tt LCRS} to maintain positional accuracy \citep{Yoder:1995}.

Topographic data from the LRO’s Lunar Orbiter Laser Altimeter (LOLA) provide vertical accuracy to approximately 1 meter and horizontal resolution down to 10 meters. These measurements facilitate the identification of stable reference points, especially in polar regions like Shackleton Crater, where libration effects are minimized. The lunar geoid, as modeled from GRAIL data, provides an equipotential reference surface with altitude variations up to 100 meters due to mascons, establishing a baseline for elevation in the {\tt LCRS} \citep{Smith-etal:2010}.

LLR has provided key measurements for understanding the Moon's rotational and tidal responses. By precisely measturing the distance between Earth and retroreflectors on the lunar surface, LLR enables the determination of the Moon’s rotational and orbital dynamics to sub-centimeter accuracy. These measurements help constrain the tidal Love numbers $k_2$ and $k_3$, refine our understanding of the Moon's interior elasticity, and reveal core-mantle interactions. Furthermore, LLR data aids in tracking librational motions and Earth-induced perturbations, essential for modeling positional stability in the {\tt LCRS}. The continuous analysis of LLR data contributes to a more detailed understanding of tidal deformation and gravitational harmonics, both critical for establishing a robust, geospatially stable reference framework within the {\tt LCRS} \citep{Dickey-etal:1994, Williams-etal:2014}.

Recent analyses of combined GRAIL, LRO and LLR data indicate a solid inner core within the Moon, with an estimated radius of \((258 \pm 40)\) km and a density of \((7,822 \pm 1,615)\) kg/m\(^3\), along with evidence of a past mantle overturn event that redistributed internal material \citep{Briaud-etal:2023}. The presence of a dense inner core impacts the Moon’s mass distribution, modifying its gravitational potential and requiring updates to existing gravitational field models. Precise gravitational modeling of the Moon’s internal structure, including the core, is fundamental for the {\tt LCRS}, as mass anomalies associated with core density affect the accuracy of positional stability, particularly in regions with mascon (mass concentration) influences.

These high-precision gravitational, topographic, and temporal datasets form the technical foundation required to develop the {\tt LCRS}, enabling accurate positioning, navigation, and infrastructure planning for future lunar operations.

In addition to these foundational elements, transforming time between the {\tt GCRS}, {\tt BCRS}, and {\tt LCRS} requires precise relativistic corrections to account for differences in gravitational potentials and relative motion between these coordinate systems. In what follows, we define the {\tt LCRS} framework and establish these important corrections.

\subsection{Defining the LCRS}
\label{sec:LCRS}

As we saw in Sec.~\ref{sec:time-E}, the constant rates for the transformations of this chain are $L_{\tt G}/(1  -  L_{\tt G})$, $L_{\tt C}/(1  -  L_{\tt B})$, and $ - L_{\tt B}$. The sum of the three is zero giving $L_{\tt C}/(1  -  L_{\tt B}) = L_{\tt B}  -  L_{\tt G}/(1  -  L_{\tt G})$, so $L_{\tt C} = L_{\tt B} (1  -  L_{\tt B})  -  L_{\tt G} (1  -  L_{\tt B})/(1  -  L_{\tt G}) \approx L_{\tt B}  -  L_{\tt G}$. 

Ultimately, to this chain we want to add ${\tt TDB} \rightarrow {\tt TCB}$ $\rightarrow$ Luni-centric Coordinate Time ({\tt TCL})  in the Luni-centric Celestial Reference System ({\tt LCRS})$\rightarrow$ Lunar Time ({\tt TL}) at the surface of the Moon, all  with appropriate rates for the transformations. Below, we will establish the  relations appropriate for this purpose, accurate to $\sim1 \times 10^{-16}$. 

Given the fact that  the {\tt BCRS} (\ref{eq:metric-CB}) is a common reference system for the solar system, to define the {\tt LCRS}, we will use the same approach as we used to define {\tt GCRS} (see Sec.~\ref{sec:B-GCRS}.) Accordingly, similar to the structure of the {\tt GCRS} metric  (\ref{eq:metric-C}), the {\tt LCRS} with coordinates ($\cal T$ = $ \tt TCL$, $ \vec {\cal{X}}$) is defined by the metric that may be given in the form, as below:
{}
\begin{equation}
ds^2_{\tt LCRS} = \Big\{1 - 2c^{-2}\Big(U_{\tt M}({\cal T}, \vec  {\cal X}) + U^\star_{\rm tid}({\cal T}, \vec  {\cal X})\Big)   +{\cal O}(c^{-4})\Big\}c^2d{\cal T}^2 - \Big\{1 + 2c^{-2}\Big(U_{\tt M}({\cal T}, \vec {\cal X}) + U^\star_{\rm tid}({\cal T}, \vec {\cal X})\Big)+{\cal O}(c^{-4})\Big\} d\vec {\cal X}^2,
\label{eq:metric-CM}
\end{equation}
where  $U_{\tt M}({\cal T}, \vec {\cal X})$ is the Newtonian gravitational potential  of the isolated Moon and $U^\star_{\rm tid}({\cal T}, \vec {\cal X})$ is  the tidal potential produced by all other solar system bodies (excluding Moon). The omitted terms are due to $U_{\tt M}^2$ that may contribute up to  $\sim 8.43\times 10^{-19}$, negligible for our purposes.

Finally, similar to the coordinate transformations between  {\tt GCRS}  and  {\tt BCRS} from (\ref{eq:coord-tr-T1-rec})--(\ref{eq:coord-tr-Xrec}),  the time and position transformations between the {\tt LCRS} ($\cal T$ = $\tt TCL$, $\vec {\cal X}$) and the {\tt BCRS} ($t$ = $\tt TCB$, $\vec x$) that may be given as follows:
{}
\begin{eqnarray}
{\cal T} &=& t-c^{-2}\Big\{\int^{t}_{t_{0}}\Big( {\textstyle\frac{1}{2}}v_{\tt M}^2+\sum_{{\tt B}\not= {\tt M}} U_{\tt B}  \Big)dt + (\vec v_{\tt M} \cdot \vec r_{\tt M}) \Big\} +
 {\cal O}\big(1.22\times 10^{-16}\big)(t-t_0),
 \label{eq:coord-tr-T1-recM}\\[4pt]
\vec {\cal X} &=& \Big(1+  c^{-2}\sum_{{\tt B}\not= {\tt M}} U_{\tt B}\Big)\vec r_{\tt M}+ c^{-2}{\textstyle\frac{1}{2}}( \vec v_{\tt M} \cdot\vec r_{\tt M})\vec v_{\tt M}  + {\cal O}\big(1.00\times 10^{-7}~{\rm m}\big),
\label{eq:coord-tr-XrecM}
\end{eqnarray}
where $\vec r_{\tt M} \equiv \vec x - \vec x_{\tt M}(t)$ with $\vec x_{\tt M}$  and $\vec v_{\tt M}=d\vec x_{\tt M}/dt$ being the Moon's position and velocity vectors  in the {\tt BCRS}. The  uncertainty in (\ref{eq:coord-tr-T1-recM}) came from the omitted ${\cal O}(c^{-4})$ terms that behave $\sim c^{-4}\big\{-\frac{1}{8}v_{\tt M}^2-\frac{3}{2}v^2_{\tt M}G M_\odot/r_{\tt M}+\frac{1}{2}(G M_\odot/r_{\tt M})^2\big\}\lesssim -1.22\times 10^{-16}= -10.50$~ps/d. The uncertainty in  (\ref{eq:coord-tr-XrecM}) is set by the omitted acceleration-depended terms that on the lunar surface may contribute up to $c^{-2}\big((\vec a_{\tt M}\cdot \vec r_{\tt M})\vec r_{\tt M}- \frac{1}{2}r^2_{\tt M}\vec a_{\tt M}\big)\simeq 1.50\times 10^{-7}$~m.  If evaluated at the Earth's distance, this term contributes only  $\sim 7.09 \times 10^{-3} \, \text{m}$, which is negligible for our purposes. The inverse transformations to (\ref{eq:coord-tr-T1-recM})--(\ref{eq:coord-tr-XrecM}) are obtained by moving the $c^{-2}$-terms on the other size of the equations with appropriate sign changes. 

Below, we will use  (\ref{eq:metric-CM}) and (\ref{eq:coord-tr-T1-recM})--(\ref{eq:coord-tr-XrecM})  to establish the new relativistic time scales for the Moon.

\subsection{Transformation {\tt TL }$\leftrightarrow$ {\tt TDB} }
\label{sec:tm-tdb}

Similar to (\ref{eq:LG}), the time transformation from the  Lunar Time ({\tt TL}) at or near the Moon’s surface to Luni-centric Coordinate Time  ({\tt TCL}) involves a rate change\footnote{
Defining a time scale {\tt TL} for the lunar system, similar to {\tt TT} in the geocentric system, is possible but not essential. Historically, time scales like {\tt TT} and {\tt TDB} were widely used on Earth before any formal framework was established. In the late 20th century, the IAU introduced {\tt TCG} and {\tt TCB} as official time scales, with {\tt TT} and {\tt TDB} derived and retained for practical continuity. For the Moon, however, no tradition or independent need for a dedicated time scale is yet exists, making {\tt TL} optional rather than necessary.} 
{}
\begin{align}
\Big<\frac{	d{\tt TL}}{d{\tt TCL}}\Big> = 1 - \frac{1}{c^2}\big<W_{\tt gM}\big> \equiv 1  -  L_{\tt L},
\qquad {\rm or} \qquad\frac{d{\tt TCL}}{d{\tt TL}} = \frac{1}{1  -  L_{\tt L}} = 1 +  \frac{L_{\tt L}}{1  -  L_{\tt L}},
\label{eq:(26)}
\end{align}
where $W_{\tt gM}$ is the combined gravitational and rotational potential on the selenoid (i.e., geoid for the Moon). 

There is significant uncertainty in determining \( W_{\tt gM} \), the reference level surface of the selenopotential. Reported \( W_{\tt gM} \) values vary widely, from \(2825390~\mathrm{m}^2\mathrm{s}^{-2} \) \cite{Bursa-Sima:1980}, derived from gravity measurements at the Apollo 12 landing site, to \( 2821713.3~\mathrm{m}^2\mathrm{s}^{-2} \) \cite{Martinec-Pec:1988}, based on a lunar gravity model \cite{Ferrari-etal:1980} utilizing Doppler tracking data from Lunar Orbiter 4 and LLR data, adjusted for lunar topography. More recently, \( W_{\tt gM} = (2822336.927 \pm 23)~\mathrm{m}^2\mathrm{s}^{-2} \) \cite{Ardalan-Karimi:2014} was determined using pre-GRAIL global gravity models (GGMs), incorporating topographic bias corrections on geoidal heights.

Currently, data from the GRAIL mission \cite{Zuber-etal:2013,Lemoine-etal:2013} provides a good basis for establishing a high-resolution lunar geoid\footnote{For details, see \url{https://pgda.gsfc.nasa.gov/products/50}.} and refining the constant $L_{\tt L}$. The LOLA dataset, which represents the Moon's shape and elevation, requires geoid subtraction to achieve accurate referencing. Addressing mascon-induced anomalies in the geoid enables the recalibration of $L_{\tt L}$, ensuring it better reflects the Moon's equipotential surface.

While the slopes of the lunar geoid are less pronounced compared to Mars, significant shape-topography slope discrepancies persist near the boundaries of nearside mascons. Utilizing GRAIL data and the latest GRGM1200A lunar gravity model\footnote{Details on the GRGM1200A model are available at \url{https://pgda.gsfc.nasa.gov/products/50}.} \cite{Goossens-etal:2016}, which incorporates adjustments based on LOLA topography measurements, the geoid potential value $W_{\tt gM}$ reported by \cite{Martinec-Pec:1988} has been further validated \cite{Cziraki-Timar:2023}. Consequently, adopting $W_{\tt gM}=2821713.3~\mathrm{m}^2\mathrm{s}^{-2}$ from \cite{Martinec-Pec:1988}, the constant $L_{\tt L}$ can be determined as: $L_{\tt L} = 3.139\,579\,5 \times 10^{-11} \simeq 2.7126~\mu\mathrm{s}/\mathrm{d}.$

Alternatively, in analogy to $L_{\tt G}$ from (\ref{eq:(5)}), the constant $L_{\tt L}$ can be modeled relying on the lunar gravity and spin:
{}
\begin{align}
L_{\tt L} \equiv \frac{1}{c^2}\Big<W_{\tt gM}\Big>=\frac{1}{c^2}\Big<U_{\tt M} + {\textstyle\frac{1}{2}} [ \vec \omega_{\tt M}\times \vec R_{\tt M}]^2\Big>\simeq \frac{1}{c^2}\Big\{ \frac{GM_{\tt M}}{R_{\tt MQ}} \big( 1 + {\textstyle\frac{1}{2}}J_{2\tt M}\big) + {\textstyle\frac{1}{2}} R_{\tt MQ}^2  \omega_{\tt M}^2 \Big\}+{\cal O}(2.11 \times 10^{-15}).
\label{eq:(29)}
\end{align}

The adopted lunar reference radius for gravity is $R_{\tt MQ} = 1738.0$~km, which is larger than the mean radius of $R_{\tt M} = 1737.1513$~km \cite{Smith-etal:2017}, the lunar gravitational constant $GM_{\tt M} = 4902.800118~{\rm km}^3/{\rm s}^2$ (DE440, \cite{Park-etal:2021}), gravity harmonic $J_{2 \tt M}$ is $2.033\times 10^{-4}$ \cite{Williams-etal:2014}, and $\omega_{\tt M} = 2\pi/(27.321\,661~{\rm d} \times 86400~{\rm s/d})=2.6616996\times 10^{-6}~{\rm s}^{-1}$. The value of $L_{\tt L}$, with the larger radius, $R_{\tt MQ} $, is then estimated to be $L_{\tt L}=
3.139\,054\,1 \times 10^{-11}\simeq 2.7121 \,\mu$s/d. Also, the error bound in (\ref{eq:(29)}) is set by the omitted term with the tesseral harmonics $C_{22}=2.242\,615\times 10^{-5}$ of the Moon's gravity field \cite{Konopliv-etal:2013}. Note that, if the smaller value  for the lunar radius $R_{\tt M}$ is used in (\ref{eq:(29)}), the result is $L_{\tt L}=
3.140\,587\,7\times 10^{-11}\simeq 
2.7135\,\mu$s/d.

The tidal terms derived from (\ref{eq:metric-CM}) may contribute to $L_{\tt L}$ at levels of up to $\sim 1.33 \times 10^{-18}$ for the Sun and $2.36 \times 10^{-16}$ for the Earth. While the solar tide lies beyond the stated accuracy threshold, the Earth tide is at the cusp of significance \cite{Turyshev-Toth:2023-grav-phase}. The Earth's tidal effects induce both spatial and temporal variations in lunar gravity, necessitating careful consideration in the definition of the selenoid. At present, both tidal terms are omitted in (\ref{eq:(29)}).

One can further improve the model (\ref{eq:(29)}) by taking into account the values of the lowest degree and order  harmonic coefficients,    
$J_{2\tt M}=2.033\times 10^{-4}$,
$C_{21}=0$, 
$S_{21}=0$, 
$C_{22}=2.242\,615\times 10^{-5}$,  
$S_{22}=0$,  
$J_{3\tt M}=8.459\,703\times 10^{-6}$,  
$C_{31}=2.848\,074\times 10^{-5}$, 
$S_{31}=5.891\,555\times 10^{-6}$,
$C_{32}=4.840\,499\times 10^{-6}$, 
$S_{32}=1.666\,142\times 10^{-6}$,
$C_{33}=1.711\,660\times 10^{-6}$, 
$S_{33}=-2.474\,276\times 10^{-7}$, and 
$J_{4\tt M}=-5.901 \times 10^{-6}$,   see \cite{Park-etal:2021, Williams-etal:2014}. As a result, the analytical model for $L_{\tt L}$ will acquire longitude and colatitude angular dependence. Including  terms only up to the third degree and order, due to the angular dependence \cite{Turyshev-Toth:2023-grav-phase}, the result will  vary between  $3.138\,859\,0 \times 10^{-11}\simeq 2.7120 \,\mu$s/d and $3.139\,318\,8 \times 10^{-11}\simeq 2.7124 \,\mu$s/d, with accuracy  limited by the omitted $J_{4\tt M}$ that contributes the error on the order of $6.94\times 10^{-17}\simeq 6~{\rm ps/d}$.

In addition to the angular dependence of the selenoid's model, the Moon's elastic nature and its tidal response induce temporal variability in its gravity harmonics, further complicating the analytical characterization of the selenoid, especially if higher spatial resolution is required. This challenge is analogous to the complexities associated with the Earth's geoid and the constant \( L_{\tt G} \), as discussed in Sec.~\ref{sec:GCRS}. These difficulties ultimately led the IAU to adopt \( L_{\tt G} \) as a defining constant in its 2000 Resolution B1.9 \cite{Soffel-etal:2003, Kaplan:2005}. A similar approach could be considered for defining \( L_{\tt L} \).  Until such refinements are achieved, we continue to rely on (\ref{eq:(29)}) as an estimate for the value of $L_{\tt L}$, shown in Table \ref{tab:1}.

The constant  $L_{\tt L}$ allows us to establish the scaling of  coordinates and mass factors to maintain the invariance of the speed of light and the equations of motion in the {\tt LCRS},  for the transformation from {\tt TCL} to {\tt TL}. Similarly to (\ref{eq:interv}), this transformation, which includes the scaling of temporal and spatial coordinates and mass factors, ensures the invariance of the metric (up to a constant factor) and has the form:  
{}
\begin{equation}
(ds^2)_{\tt TL} = (1- L_{\tt L})^2ds^2_{\tt TCL} ,
\label{eq:interv-M}
\end{equation}
where $(ds^2)_{\tt TL}$ maintains the same form in terms of {\tt TL}, $\vec {\cal X}_{\tt TL}$, and $(G M )_{\tt TL}$ as  (\ref{eq:metric-CM}) does in terms of $\cal T$, $\vec {\cal X}$, and $(GM)_{\tt TCL}$. 

As a result, instead of using coordinate time ${\cal T} = {\tt TCL}$, spatial coordinates $\vec{\cal X}$, and mass factors $(G M)_{\tt TCL}$ related to the ({\tt LCRS}), we will use the scaling for the relevant quantities in the Lunar Surface Coordinate Reference System ({\tt LSCRS}). To establish these relations, we integrate (\ref{eq:(26)}) from ${\tt T_{L0}}$ to ${\tt TCL}$, deriving the connection between the two time scales. Additionally, the spatial coordinates and mass factors are adjusted in accordance with (\ref{eq:interv-M}), resulting in:
{}
\begin{equation}
{\tt TL} = {\tt TCL}- L_{\tt L}({\tt TCL}-{\tt T}_{\tt L0}), \qquad
\vec {\cal X}_{\tt TL} = (1-L_{\tt L})\vec {\cal X}_{\tt TCL}, \qquad
(GM)_{\tt TL} = (1-L_{\tt L})(GM)_{\tt TCL},
\label{eq:LSCRS}
\end{equation}
where $ {\tt T_{L0}} $  is the initial lunar time, which, for now, we will use unspecified. 

Similar to (\ref{eq:synch}), and using (\ref{eq:metric-CM}) and (\ref{eq:(26)}), we express the proper time of an ideal clock located near the selenoid (e.g., on the surface of the Moon or in lunar orbit, while moving with velocity ${\vec v}_{\tt C0}$ in {\tt LCRS}), \( \tau^\star_{\tt C} \), relative to {\tt TL}, as follows:
{}
\begin{equation}
\frac{d\tau^\star_{\tt C}}{d{\tt TL}}=\frac{d\tau^\star_{\tt C}}{d{\tt TCL}}\frac{d{\tt TCL}}{d{\tt TL}}\simeq 1+L_{\tt L}-\frac{1}{c^2}\Big\{{\textstyle\frac{1}{2}}{\vec v}^2_{\tt C0}
+ U_{\tt M}({\vec {\cal X}}_{\tt C})+ U^\star_{\rm tid}(\vec  {\cal X}_{\tt C})\Big\}_{\tt TL}+{\cal O}(c^{-4}).
\label{eq:synch-M}
\end{equation}

The transformation from {\tt TCL} to {\tt TCB} is analogous to Eq.~(\ref{eq:(12)}) and from (\ref{eq:coord-tr-T1-recM}) is determined to be
{}
\begin{align}
{\tt TCL} -{\tt TCB}  = -\frac{1}{c^2}\Big\{\int \Big( {\textstyle\frac{1}{2}}v_{\tt M}^2 +  \sum_{{\tt B}\not={\tt M}}U_{\tt B} \Big)d{\tt TCB}+\big(\vec v_{\tt M} \cdot \vec {r_{\tt M}}_{\tt TCB}\big)\Big\}  +{\cal O}(c^{-4}),
\label{eq:(30-n)}
\end{align}
where $\vec v_{\tt M}$ is the solar system barycentric velocity vector of the Moon, $v_{\tt M}$ is its scalar, $U_{\tt M}$ is the external potential at the Moon’s center, and $\vec r_{\tt M}=\vec x-\vec x_{\tt M}$ is the {\tt BCRS} vector from the center of the Moon to the surface site. The potential and kinetic energy use the Moon centered reference frame.  The dot product annually reaches $\pm0.58~\mu$s with smaller variations  of $\pm 21$~ps at Moon's sidereal period of $t_{\tt M}=27.32166~{\rm d}$.

In analogy to (\ref{eq:(12inv)}), the time transformation  {\tt TCB} $\rightarrow$ {\tt TCL}, which is inverse to (\ref{eq:(30-n)}), has the form
{}
\begin{align}
{\tt TCB}  -{\tt TCL} = \frac{1}{c^2}\Big\{\int \Big( {\textstyle\frac{1}{2}}v_{\tt M}^2 +  \sum_{{\tt B}\not={\tt M}}U_{\tt B} \Big)d{\tt TCL} +\big(\vec v_{\tt M} \cdot \vec {{\cal X}}_{\tt TCL}\big)\Big\}  +{\cal O}(c^{-4}),
\label{eq:(30-n-inv)}
\end{align}
where the subscript  $\{...\}_{\tt TCL}$ is used to identify a ${\tt TCL}$-compatible quantity. 

Similar to (\ref{eq:coord-tr-QQ}), ${\tt TCB} - {\tt TCL}$ from (\ref{eq:(30-n)}) and (\ref{eq:(30-n-inv)})  has a mean rate given by constant $L_{\tt H}$  
{}
\begin{eqnarray}
\frac{1}{c^2}\Big({\textstyle\frac{1}{2}}v_{\tt M}^2 +  \sum_{{\tt B}\not={\tt M}}U_{\tt B} \Big)=L_{\tt H}+
\dot P_{\tt H}(t)+{\cal O}(c^{-4}), \qquad\qquad 
\frac{1}{c^2}\Big<{\textstyle\frac{1}{2}}v_{\tt M}^2
+  \sum_{{\tt B}\not={\tt M}}U_{\tt B} \Big>=L_{\tt H},
 \label{eq:coord-tr-QQM}
\end{eqnarray}
where constant $L_{\tt H}$ results from the long-time averaging of the Moon's total orbital energy in the {\tt BCRS} determined as  $L_{\tt H}=1.482\,536\,24 \times 10^{-8} \approx 1.280\,913\,2$~ms/d, and $P_{\tt H}(t)$ represents a series of small periodic terms. The quantity $P_{\tt H}(t)$ can be evaluated semi-analytically in the same manner as the time-series $P_{\tt }(t)$ for the Earth, e.g. \cite{Fairhead-Bretagnon:1990,Fukushima:1995,Irwin-Fukushima:1999,Fukushima:2010}. 

Note that the (positive) external potential $\sum_{{\tt B}\not={\tt M}}U_{\tt B}$  at the Moon’s center from the Sun, Earth, and planets in (\ref{eq:coord-tr-QQM}) is analogous to the external potential  $\sum_{{\tt B}\not={\tt E}}U_{\tt B}$ at the geocenter from the Sun, Moon, and planets, given by (\ref{eq:coord-tr-QQ}). The Earth and Moon are accelerated by an average acceleration of $5.9~{\rm mm/s}^2$ by the Sun, and the Moon is accelerated by an average of $2.7~{\rm  mm/s}^2$ by the Earth. However, the potential of the Sun is three orders of magnitude larger than the potential of the Earth at the Moon, and five orders of magnitude larger than the potential of the Moon at the Earth.

Eq.~(\ref{eq:(30-n)}) together with (\ref{eq:coord-tr-QQM}) gives the mean rate between ${\tt TCL}$ and ${\tt TCB}$:
{}
\begin{equation}
\Big<\frac{d{\tt TCL}}{d{\tt TCB}}\Big> =1-L_{\tt H}.
\label{eq:constTCGTCBH}
\end{equation}

To express  ${\tt TCB}$ via ${\tt TL}$, we need another constant that we call $L_{\tt M}$, which determines the rate  between ${\tt TL}$  and ${\tt TCB}$ and, similarly to (\ref{eq:LB}),  may be formally introduced as 
{}
\begin{equation}
\Big<\frac{d{\tt TL}}{d{\tt TCB}}\Big> =1-L_{\tt M}.
\label{eq:LM}
\end{equation}
In the analogy with (\ref{eq:(nonN3)}), which determined the constant $L_{\tt B}$ shown in Table~\ref{tab:1}, we compute the rate $L_{\tt M}$ as below 
{}
\begin{align}
L_{\tt M} = \frac{1}{c^2}\Big\{ GM_{\tt S} \Big<\frac{1}{r_{\tt MS}}\Big> + GM_{\tt E} \Big<\frac{1}{r_{\tt ME}}\Big> + \big<U_{\tt MP}\big> + \big<{\textstyle\frac{1}{2}}v_{\tt M}^2\big>+ \frac{GM_{\tt M}}{R_{\tt MQ}} ( 1 + {\textstyle\frac{1}{2}}J_{\tt 2M}) + {\textstyle\frac{1}{2}} R_{\tt MQ}^2 \omega_{\tt M}^2 \Big\} +{\cal O}(2.11 \times 10^{-15}),
\label{eq:(nonN4)}
\end{align}
with the error bound is set by the omitted term with the tesseral harmonics $C_{22}$ of the lunar gravity field, as in (\ref{eq:(29)}), with the cumulative effect of the higher harmonics terms omitted (\ref{eq:(29)}) being on the same order, e.g., $2\times 10^{-15}$.  Thus, the analytical definition of $L_{\tt M}$ with an accuracy below $10^{-15}$ encounters similar technical challenges—such as spatial and temporal variability at higher degrees and orders of spherical harmonics—as those discussed above for $L_{\tt L}$. This may necessitate declaring $L_{\tt M}$ as a defining constant for the {\tt LCRS}, analogous to the treatment of $L_{\tt B}$ in the {\tt GCRS}. 

Table~\ref{tab:1} summarizes various constants in the rate equations for Earth and Moon. The $GM$ values for the Sun, Earth, Moon, and planets were taken from DE440 \cite{Park-etal:2021}. These will not exactly match the values used for the IAU value of ${L_{\tt B}}$. 
The potentials of the planets other than Earth, $\big<U_{\tt MP}\big>$, are computed and discussed in Appendix~\ref{sec:planet-GR}. Using these values the constant was found to be $L_{\tt M} =
1.485\,675\,290 \times 10^{-8}\approx 1.283\,62$~ms/d. 

We can use the chain of time derivatives, to establish the relationships between the constants $L_{\tt M}, L_{\tt L}$ and $L_{\tt H}$, similar to (\ref{eq:constLBLCLG}). Following this approach, with the help of (\ref{eq:(26)}), (\ref{eq:constTCGTCBH}) and (\ref{eq:LM}), we have the following expression  
{}
\begin{equation}
\Big<\frac{d{\tt TL}}{d{\tt TCB}}\Big>=\Big(\frac{d{\tt TL}}{d{\tt TCL}}\Big)\Big<\frac{d{\tt TCL}}{d{\tt TCB}}\Big> \qquad \Rightarrow \qquad (1-L_{\tt M})=(1-L_{\tt L})(1-L_{\tt H}),
\label{eq:consTM}
\end{equation}
from which the constant $L_{\tt M}$ is determined as $L_{\tt M} \simeq L_{\tt L}+L_{\tt H}-L_{\tt L}L_{\tt H}=
1.485\,675\,294\times 10^{-8}\approx 1.283\,62$~ms/d.

Results  (\ref{eq:(26)}), (\ref{eq:LSCRS}), allow us to express $d{\tt TCL}=(d{\tt TCL}/d{\tt TL})d{\tt TL}=d{\tt TL}/(1-L_{\tt L})$ and $\vec {\cal X}_{\tt TCL} = \vec {\cal X}_{\tt TC}/(1-L_{\tt L})$, correspondingly. With this, we use (\ref{eq:coord-tr-QQM}) and integrate  (\ref{eq:(30-n-inv)}) from ${\tt T}_{\tt L0} $ to {\tt TL} and present it as a function of {\tt TL}, as below
{}
\begin{equation}
{\tt TCB} - {\tt TCL} = \frac{L_{\tt H}}{1-L_{\tt L}}
 ({\tt TL}-{\tt T}_{\tt L0})+P_{\tt H}({\tt TL})-P_{\tt H}({\tt T}_{\tt L0}) +\frac{1}{c^2} (\vec v_{\tt M} \cdot \vec {\cal X}_{\tt TL}) +{\cal O}(c^{-4}),
\label{eq:(12)TM}
\end{equation}
which comprises a secular term with the rate of $\propto L_{\tt H}/(1-L_{\tt L})=
1.482\,536\,240 \times 10^{-8}=1.281$~ms/d and a series of periodic terms $\propto P_{\tt H}({\tt TL})$. 

To develop the  $({\tt TCB} - {\tt TL})$ transformation, we 
work with the following chain $({\tt TCB} - {\tt TL})=({\tt TCB} - {\tt TCL})+({\tt TCL} - {\tt TL})$
and again use (\ref{eq:(26)}),  (\ref{eq:LSCRS}),  (\ref{eq:(30-n-inv)}), to derive the following result: 
{}
\begin{equation}
{\tt TCB} - {\tt TL} = \frac{L_{\tt L}}{1-L_{\tt L}}({\tt TL}-{\tt T}_{\tt L0})
+
\frac{1}{c^2}\Big\{\int_{\tt T_{\tt L0}}^{\tt TL} \Big( {\textstyle\frac{1}{2}}v_{\tt M}^2 +  \sum_{{\tt B}\not={\tt M}}U_{\tt B} \Big)d{\tt TL} + 
(\vec v_{\tt M}\cdot \vec {\cal X}_{\tt TL})\Big\}  +{\cal O}(c^{-4}).
\label{eq:(12)TMTM-int}
\end{equation}
The same transformation, but evaluating the integral over $d{\tt TL}$ with the help of expression (\ref{eq:coord-tr-QQM}) results in
{}
\begin{equation}
{\tt TCB} - {\tt TL} = \frac{L_{\tt M}}{1-L_{\tt L}}
({\tt TL}-{\tt T}_{\tt L0})+P_{\tt H}({\tt TL})-P_{\tt H}({\tt T}_{\tt L0})+\frac{1}{c^2} (\vec v_{\tt M} \cdot \vec {\cal X}_{\tt TL})  +{\cal O}(c^{-4}),
\label{eq:(12)TMTM} 
\end{equation}
which, compared to (\ref{eq:(12)TM}), has a sightly different secular drift with the rate  of $\propto L_{\tt M}/(1-L_{\tt L})=1.485\,675\,290 \times 10^{-8}=1.284$~ms/d and small periodic terms $\propto P_{\tt H}({\tt TL})$. 

Expressions  (\ref{eq:(12)TMTM-int}) and (\ref{eq:(12)TMTM}) allow us to develop $({\tt TDB}  -  {\tt TL})$ as a function of {\tt TL}.  For this purpose, using (\ref{eq:TDBC}), (\ref{eq:consTM}), and (\ref{eq:(12)TMTM-int}), we express  $({\tt TDB} - {\tt TL})$ as a function of {\tt TL} as below 
{}
\begin{eqnarray}
{\tt TDB}  -  {\tt TL} &=&   {\tt TDB}_0- \frac{L_{\tt B}-L_{\tt L}}{1-L_{\tt L}} \,\big({\tt TL}-{\tt T}_{\tt L0}\big)  +L_{\tt B}({\tt T}_0-{\tt T}_{\tt LO})+
\frac{1}{c^2}\Big\{\int_{\tt T_{\tt L0}}^{\tt TL} \Big( {\textstyle\frac{1}{2}}v_{\tt M}^2 +  \sum_{{\tt B}\not={\tt M}}U_{\tt B} \Big) d{\tt TL} +
(\vec v_{\tt M}\cdot \vec {\cal X}_{\tt TL})\Big\}+{\cal O}(c^{-4}).~~~
\label{eq:(nonN2-int)}
\end{eqnarray}

Evaluating integral in (\ref{eq:(nonN2-int)}) with the help of (\ref{eq:coord-tr-QQM}), yields in the following expression:  
{}
\begin{align}
{\tt TDB}  -  {\tt TL} =  {\tt TDB}_0 - \frac{L_{\tt B}-L_{\tt M}}{1-L_{\tt L}}\big({\tt TL} -{\tt T}_{\tt L0}\big) +L_{\tt B}({\tt T}_0-{\tt T}_{\tt LO})+ 
P_{\tt H}({\tt TL})-P_{\tt H}({\tt T}_{\tt L0}) + \frac{1}{c^2}(\vec v_{\tt M}\cdot \vec {\cal X}_{\tt TL})
+{\cal O}(c^{-4}).
\label{eq:(nonN2)}
\end{align}
Thus, there is a secular drift  between {\tt TDB} and {\tt TL} at a rate of   $(L_{\tt B}-L_{\tt M})/(1-L_{\tt L})=
6.484\,440\,313\times 10^{-10}=56.0256~\mu$s/d.

It is useful to express the difference $({\tt TDB} - {\tt {\tt TL}})$ as a function of {\tt TDB}. This can be done by using (\ref{eq:TDBC}), (\ref{eq:LSCRS}), (\ref{eq:(30-n)}), and (\ref{eq:consTM}), yielding result  below
{}
\begin{eqnarray}
{\tt TDB}  -  {\tt TL} &= &
\frac{1-L_{\tt L}}{1-L_{\tt B}}\,
{\tt TDB}_0
 -\frac{L_{\tt B}-L_{\tt L}}{1-L_{\tt B}} \big( {\tt TDB}-{\tt T}_0\big)  +L_{\tt L}({\tt T}_0-{\tt T}_{\tt LO})+ \nonumber\\
 &+&\frac{1-L_{\tt L}}{1-L_{\tt B}}\frac{1}{c^2}\Big\{\int_{{\tt T_0+TDB_0}}^{\tt TDB}\Big( {\textstyle\frac{1}{2}}v_{\tt M}^2 +  \sum_{{\tt B}\not={\tt M}}U_{\tt B} \Big) d{}\tt TDB +
 (\vec v_{\tt M}\cdot \vec {r_{\tt M}}_{\tt TDB})\Big\}+{\cal O}(c^{-4}).
\label{eq:(nonN2-in+)}
\end{eqnarray}

Finally,  evaluating the integral in (\ref{eq:(nonN2-in+)}) with the help of (\ref{eq:coord-tr-QQM}), we derive the following result:
{}
\begin{eqnarray}
{\tt TDB}  -  {\tt TL} &=&   \frac{1-L_{\tt M}}{1-L_{\tt B}}\,
{\tt TDB}_0 - \frac{L_{\tt B}-L_{\tt M}}{1-L_{\tt B}} \big({\tt TDB}-{\tt T}_0\big)  +L_{\tt L}({\tt T}_0-{\tt T}_{\tt LO})+\nonumber\\&+& 
P_{\tt H}({\tt TDB}) -P_{\tt H}({\tt T}_0+{\tt TDB}_0)+ \frac{1}{c^2}(\vec v_{\tt M}\cdot \vec {r_{\tt M}}_{\tt TDB})
+{\cal O}(c^{-4}).~~
\label{eq:(nonN2+)}
\end{eqnarray}

As seen from (\ref{eq:(nonN2+)}), there is a rate difference between  ${\tt TL}$ and  ${\tt TDB}$, that is given by the combination of the constants 
 $ (L_{\tt B}-L_{\tt M})/(1-L_{\tt B})=
6.484\,440\,414 \times 10^{-11}\simeq 56.0256~\mu$s/d, with {\tt TL} running faster than {\tt TDB}. In addition, there is also a series of small periodic terms $\propto P_{\tt H}({\tt TDB})$ and the term that depends on the lunar surface position $(\vec v_{\tt M}\cdot \vec {r_{\tt M}}_{\tt TDB})/c^2$.
 
\subsection{Position transformations}
\label{sec:pos-M}

Now we need to develop position transformation for the Moon. For that, similar to introducing (\ref{eq:coord-tr-Xrec=}), we consider $\vec {\cal X}_{\tt TCL}$ to be a position vector of  a site on the surface of the Moon as observed in the {\tt LCRS} and  ${\vec r_{\tt M}}_{\tt TCB}$  being the same vector but as seen from the {\tt BCRS}.  We need to express the relationship between $\vec {\cal X}_{\tt TCL}$ and ${\vec r_{\tt M}}_{\tt TCB}$  in terms of the {\tt TL} and {\tt TDB} timescales. For that, we use (\ref{eq:LSCRS}) to express the $\vec {\cal X}_{\tt TCL}$  via ${\tt TL} $ and (\ref{eq:TDBC}) to express ${\vec r_{\tt M}}_{\tt TCB}$ via ${\vec r_{\tt M}}_{\tt TDB}$, which results in
{}
\begin{eqnarray}
\vec {\cal X}_{\tt TL} &=& \Big(1+ L_{\tt B} -L_{\tt L} +c^{-2} \sum_{{\tt B}\not= {\tt M}} U_{\tt B} \Big){\vec r_{\tt M}}_{\tt TDB}+ c^{-2} \frac{1}{2}\big( \vec v_{\tt M} \cdot\vec {r_{\tt M}}_{\tt TDB}\big)\vec v_{\tt M} +
 {\cal O}\big(1.00\times 10^{-7}~{\rm m}\big),~~~~
\label{eq:coord-tr-Xrec=M}
\end{eqnarray}
where $\vec r_{\tt M} \equiv \vec x - \vec x_{\tt M}(t)$  and 
similar to (\ref{eq:coord-tr-XrecM}), the error here is set by the omitted acceleration-dependent terms. Also, $(1-L_{\tt L})/(1-L_{\tt B})-1\simeq L_{\tt B}-L_{\tt L}=
1.547\,380\,65 \times 10^{-8}\approx 1.336\,94$~ms/d. Inverse transformation to that of (\ref{eq:coord-tr-Xrec=M})  is straightforward. For that, one simply needs to move the $c^{-2}$-terms  to the other side of the equation with the appropriate sign changes.

The first group of terms in (\ref{eq:coord-tr-Xrec=M}) amounts to a scale change. With $L_{\tt B}-L_{\tt L}=
1.547\,380\,65\times 10^{-8}$ and $c^{-2}\big<\sum_{\tt B\not=M}U_{\tt B}\big> = 0.988438\times 10^{-8}$ (Table~\ref{tab:1}) totaling $2.53582 \times 10^{-8}$. That amounts to  $4.41$~cm of scale change for the radius of the Moon. The annual variation is $\pm1.65\times 10^{-10}$, or $\pm0.29$~mm. Note that the Earth and Moon annual variations will affect the surface-to-surface distance along the center-to-center line with an annual variation of  $1.34~{\rm mm}|\cos l'|$ \cite{Williams-Boggs:2020}. With this, to a good approximation 
{}
\begin{eqnarray}
\vec {\cal X}_{\tt TL} &=& \Big(1+ 2.53586\times 10^{-8} + 1.65\times 10^{-10} \cos l' \Big){\vec r_{\tt M}}_{\tt TDB}+ c^{-2} \frac{1}{2}\big( \vec v_{\tt M} \cdot\vec {r_{\tt M}}_{\tt TDB}\big)\vec v_{\tt M} +
 {\cal O}\big(1.00\times 10^{-7}~{\rm m}\big),~~~~
\label{eq:coord-tr-XNN-M}
\end{eqnarray}
where the $\cos l'$ term accounts for the eccentricity of the annual Earth-Moon barycentric motion. 

The last term in (\ref{eq:coord-tr-XNN-M}) is the Lorentz contraction. For a site on the Moon's surface, the Lorentz contraction is  $8.7$~mm. Along the direction of motion, the combined scale change and Lorentz contraction is  $ 53$~mm, while perpendicular to the motion the scale change is  $ 44$~mm. Scale change and Lorentz contraction are significant for the definition of LLR retroreflector positions on the Moon (e.g., \cite{Williams-Boggs:2020}).  

\section{Implementation }
\label{sec:imp}

 There are two JPL computer programs that are affected by the relativistic corrections in this paper. One program integrates the positions of the Moon and planets plus the orientation of the Moon vs. time \cite{Park-etal:2021}. The second program calculates the round-trip time of flight of a laser beam from a ranging station on the Earth to a retroreflector on the Moon, where it bounces and returns to the station \cite{Williams-Boggs:2020}. Both of these programs use the {\tt BCRS} with {\tt TDB}  to perform the calculations. The transformations involving the {\tt TDB}-compatible spatial scale and Lorentz contraction of Moon-centered coordinates (discussed in Secs.~\ref{sec:pos-E} and \ref{sec:pos-M})  have been implemented in the JPL programs used to generate ephemerides of the Moon and planets \cite{Park-etal:2021}. The fully-relativistic {\tt LCRS} with {\tt TL} is yet to be implemented. 

In the near future, with missions expected to land on the Moon's surface by robotic Commercial Lunar Payload Services (CLPS) and crewed Artemis spacecraft, we need relativistic time transformations that convert between time scales used for solar system bodies ({\tt TDB}) and {\tt TL}. The relevant transformation $(\tt TDB-TL)$ is given by (\ref{eq:(nonN2-int)}) and (\ref{eq:(nonN2)}) as function of {\tt TL}, and by (\ref{eq:(nonN2-in+)}) and (\ref{eq:(nonN2+)}) as function of {\tt TDB}. The constant rate $ (L_{\tt B}-L_{\tt M})/(1-L_{\tt M})= 6.484\,440\,414\times 10^{-10}=56.0256~\mu$s/d, with the constants  $L_{\tt B}$ and $L_{\tt M}$ given in Table~\ref{tab:1}, represents the average drift rate caused by the velocity and potential functions involved in the definitions of the {\tt TDB} and {\tt TL} time scales. 

There are new recommended constants that are relevant for the relativistic time and position transformations on the Moon: Constant $L_{\tt L}$ (similar to $L_{\tt G}$ for the Earth) compensates for the average rate in the time transformation between the center of the Moon and the vicinity of its surface. Constant $L_{\tt H}$ (similar to $L_{\tt C}$) is the long-time average of the total Moon's orbital energy in its motion around the solar system's barycenter. Constant $L_{\tt M}$ (similar to $L_{\tt B}$) compensates the average rate in time transformation between the {\tt BCRS} and {\tt TL}.  The analogous transformations for the Earth has been addressed by many researchers. Those transformations are defined by a series of IAU resolutions. We calculated those transformations here as a check on our lunar work. Those constant rates are  presented in Table~\ref{tab:1}. 

\subsection{Expressing $({\tt TL} - {\tt TT})$  as a function of either {\tt TDB} or {\tt TT}}
\label{sec:tl-tt-tdb}

With the results obtained in this paper, we can establish the relationships between {\tt TT} and {\tt TL}. For this purpose, we may use (\ref{eq:(nonN1-TCB)}) and (\ref{eq:(nonN2-in+)}) that involve the common time {\tt TDB} time of the {\tt BCRS}.\footnote{Since (\ref{eq:(nonN1-TCB)}) and (\ref{eq:(nonN2-in+)}) are expressed within the same coordinate system, they can be combined directly to derive the desired result.} Using these expressions, we can write:
{}
\begin{eqnarray}
{\tt TL} -   {\tt TT} &=&\frac{L_{\tt G}-L_{\tt L}}{1-L_{\tt B}}\big( {\tt TDB}-{\tt T_0}-{\tt TDB}_0\big) -L_{\tt L}({\tt T}_0-{\tt T}_{\tt LO})+\nonumber\\
  &+& \frac{1}{c^2}\int_{\tt T_0+TDB_0}^{\tt TDB} \Big\{\Big( {\textstyle\frac{1}{2}}v_{\tt E}^2 +  \sum_{{\tt B}\not={\tt E}}U_{\tt B} \Big)-\Big( {\textstyle\frac{1}{2}}v_{\tt M}^2 +  \sum_{{\tt B}\not={\tt M}}U_{\tt B}\Big)\Big\} d{\tt TDB}+
 \frac{1}{c^2}\Big( (\vec v_{\tt E}\cdot \vec {r_{\tt EM}})-(\vec v_{\tt EM}\cdot \vec {r_{\tt M}}_{\tt TDB})\Big), 
\label{eq:(nonN2-in+2)}
\end{eqnarray}
accurate to ${\cal O}\big(1.22\times 10^{-16}({\tt TDB -T_0- TDB_0})\big)$. We used $(\vec v_{\tt E} \cdot \vec {r_{\tt E}})-(\vec v_{\tt M}\cdot \vec {r_{\tt M}})=(\vec v_{\tt E} \cdot \vec {r_{\tt EM}})-(\vec v_{\tt EM}\cdot \vec {r_{\tt M}})$, with ${r_{\tt M}}_{\tt TDB}$ being the {\tt TDB}-compatible positon of the lunar clock. The constants $ L_{\tt G}, L_{\tt C}, L_{\tt B}$  for the Earth and $L_{\tt L}, L_{\tt H}, L_{\tt M}$ for the Moon are given in Table~\ref{tab:1}.  The constants ${\tt T}_0({\tt MJD})={\rm MJD} 43144 + 32.184~{\rm s}$ and ${\tt TDB}_0 = -65.5~\mu$s are defining constants \cite{Kaplan:2005,Petit-Luzum:2010}. The constant ${\tt T}_{\tt L0}$ has yet to be chosen. Where $\vec r_{\tt E}$ and $\vec r_{\tt M}$ are the {\tt BCRS} position vectors of the Earth and Moon sites, correspondingly. 
Note that the largest term in (\ref{eq:(nonN2-in+2)}) that involves the constants multiplying the integrals, evaluated as   $c^{-2}\big\{ (L_{\tt B}-L_{\tt L})\big( {\textstyle\frac{1}{2}}v_{\tt M}^2 +  \sum_{{\tt B}\not={\tt M}}U_{\tt B}\big)  -(L_{\tt B}-L_{\tt G}) \big( {\textstyle\frac{1}{2}}v_{\tt E}^2 +  \sum_{{\tt B}\not={\tt E}}U_{\tt B} \big) \big\}\simeq 1.55\times 10^{-17}$, which is too small for our purposes.

As shown by (\ref{eq:(nonN2-in+2)}), there is a secular trend between  {\tt TT} and {\tt TL} with the rate of $(L_{\tt G}-L_{\tt L})/(1-L_{\tt B}) \simeq 6.655\,38\times 10^{-10}=57.5025\,\mu$s/d. However, this trend is partially reduced by the terms  with the total orbital energies of the Earth and the Moon that are present under the two integral signs. To demonstrate this, we use (\ref{eq:coord-tr-QQ}) and (\ref{eq:coord-tr-QQM}) (or directly apply results (\ref{eq:(nonN1TCB)}) and (\ref{eq:(nonN2+)})) and formally present relationship between {\tt TT} and {\tt TL}  in the following equivalent form 
{}
\begin{eqnarray}
{\tt TL} -  {\tt TT} &= & \frac{L_{\tt B}-L_{\tt M}}{1-L_{\tt B}}\,( {\tt TDB} -{\tt T_0}-{\tt TDB}_0) -L_{\tt L}({\tt T}_0-{\tt T}_{\tt LO})+ \nonumber\\
&+&
\Big(P({\tt TDB})-P({\tt T}_0+{\tt TDB}_0)\Big)-
\Big(P_{\tt H}({\tt TDB}) -P_{\tt H}({\tt T}_0+{\tt TDB}_0) \Big)
+\frac{1}{c^2} \Big((\vec v_{\tt E} \cdot \vec {r_{\tt EM}})-(\vec v_{\tt EM}\cdot \vec {r_{\tt M}}_{\tt TDB})\Big),
\label{eq:(nonN2+)2}
\end{eqnarray}
accurate to ${\cal O}\big(1.22\times 10^{-16}({\tt TDB-T_0- TDB_0})\big)$.
The quantity $P(t)$ represents a series with small periodic terms computed in \cite{Fairhead-Bretagnon:1990,Harada-Fukushima:2003} and used in  IAU Resolutions (e.g., \cite{Kaplan:2005,Petit-Luzum:2010}.) The series $P_{\tt H}(t)$ has yet to be semi-analytically computed and implemented for practical applications. 

The constant rate of the time drift between {\tt TL} and {\tt TT} in (\ref{eq:(nonN2+)2}) can be expressed as 
{}
\begin{eqnarray}
\frac{L_{\tt B}-L_{\tt M}}{1-L_{\tt B}}&=&\big(1339.6490 -1283.6235\big)~\mu{\rm s/d}=56.0256 ~\mu{\rm s/d}.~~
\label{eq:(nonrRR}
\end{eqnarray}

As a result, using  {\tt TDB} at the {\tt BCRS} as the common time, one can synchronize {\tt TT} with {\tt TL}. The two clocks will experience a drift at the rate of $ (L_{\tt B}-L_{\tt M})/(1-L_{\tt B})=
6.484\,440\,413 \times 10^{-10}=56.0256~\mu$s/d, with the clock on or near the Moon's surface running faster by that amount compared to its terrestrial analogue. Additionally, there are small periodic terms in $P_{\tt H}({\tt TDB})$ that can be evaluated semi-analytically in the manner used to compute  $P({\tt TDB})$, e.g., \cite{Fairhead-Bretagnon:1990,Fukushima:1995,Irwin-Fukushima:1999,Fukushima:2010}. The last two terms on  each line with periodic terms are site-dependent. Evaluated on the surfaces of the Earth and the Moon, they contribute  $(\vec v_{\tt E} \cdot \vec r_{\tt E})/c^2\simeq 2.11\,\mu$s and $(\vec v_{\tt M}\cdot \vec r_{\tt M})/c^2\simeq 0.60\,\mu$s with  diurnal and synodic periods, correspondingly. Smaller oscillatory terms with various periods are also present (e.g., \cite{Fairhead-Bretagnon:1990,Harada-Fukushima:2003}). 

Results (\ref{eq:(nonN2-in+2)}) and (\ref{eq:(nonN2+)2}) may also be given as a function of the terrestrial time {\tt TT}. In particular, with the help of  (\ref{eq:(nonN1)}), {\tt TL} may be given in terms of {\tt TT}. 
We also recognize that $\vec {r_{\tt M}}_{\tt TDB}=\vec {\cal X}_{\tt TT}+{\cal O}(c^{-2})$, with $\vec {\cal X}_{\tt TT}$ being the {\tt TT} compatible lunicentic position of the lunar clock. As a result,  (\ref{eq:(nonN2-in+2)}) can be expressed as  
{}
\begin{eqnarray}
{\tt TL} -   {\tt TT} &=&\frac{L_{\tt G}-L_{\tt L}}{1-L_{\tt B}}\big( {\tt TT}-{\tt T_0}\big) -L_{\tt L}({\tt T}_0-{\tt T}_{\tt LO})+\frac{1}{c^2}\int_{\tt T_0}^{\tt TT} \Big\{\Big( {\textstyle\frac{1}{2}}v_{\tt E}^2 +  \sum_{{\tt B}\not={\tt E}}U_{\tt B} \Big)-\Big( {\textstyle\frac{1}{2}}v_{\tt M}^2 +  \sum_{{\tt B}\not={\tt M}}U_{\tt B}\Big)\Big\} d{\tt TT}+\nonumber\\
  &+& \frac{1}{c^2}\Big((\vec v_{\tt E}\cdot \vec r_{\tt EM})-
(\vec v_{\tt EM}\cdot \vec {\cal X}_{\tt TT})\Big)+{\cal O}\Big(1.22\times 10^{-16} ({\tt TT -T_0})\Big).
\label{eq:(nonN2-in+2)-TT}
\end{eqnarray}

Similarly, with the help of (\ref{eq:(nonN1T)}) and (\ref{eq:(nonrRR}), result (\ref{eq:(nonN2-in+2)-TT})  may be formally presented in the following equivalent form 
{}
\begin{eqnarray}
{\tt TL} -  {\tt TT} &= &  \frac{L_{\tt B}-L_{\tt M}}{1-L_{\tt B}} \,( {\tt TT} -{\tt T_0}) -L_{\tt L}({\tt T}_0-{\tt T}_{\tt LO})+ \Big(P({\tt TT})-P({\tt T}_0)\Big)-\Big(P_{\tt H}({\tt TT}) -P_{\tt H}({\tt T}_0)\Big)+\nonumber\\
&+&
\frac{1}{c^2}\Big( (\vec v_{\tt E} \cdot \vec r_{\tt EM})-(\vec v_{\tt EM}\cdot \vec {\cal X}_{\tt TT}\big)\Big)
+{\cal O}\Big(1.22\times 10^{-16} ({\tt TT -T_0})\Big).
\label{eq:(nonN2+)2-TT}
\end{eqnarray}

As shown by (\ref{eq:(nonN2+)2-TT}), there is a secular trend between a clock on or near the surface of the Moon and a clock on or near the Earth's surface. This trend has the rate of $ (L_{\tt B}-L_{\tt M})/(1-L_{\tt B})=
6.484\,440\,414  \times 10^{-10}=56.0256~\mu$s/d, with the lunar clock running faster compared to its terrestrial counterpart. In addition, there are periodic terms represented by the series $P(t)$ and $P_{\tt H}(t)$, both of which can be computed analytically and evaluated using solar system and lunar ephemerides (e.g., \cite{Park-etal:2021}). Finally, there are also small site-dependent terms contributing at various specific periods. 

For the stated fractional accuracy of ${\cal O}(1.22\times 10^{-16})$, we may omit $L_{\tt B}$ from the denominator in the clock rate in (\ref{eq:(nonN2+)2-TT}), as it contributes only $(L_{\tt B}-L_{\tt M})L_{\tt B}\simeq 1.01\times 10^{-17}$. We retained it to indicate the structure of this quantity in case a higher accuracy expression is needed. Given that the largest term in the $P$-series has an amplitude of $A_1=1656.674564 \times 10^{-6}$~s and a frequency of $\omega_1\simeq 1.9923515 \times 10^{-7}~{\rm s}^{-1}$ \cite{Fairhead-Bretagnon:1990,Petit-Luzum:2010,Harada-Fukushima:2003}, the presence of the factor $(1-L_{\tt G})/(1-L_{\tt B})=1/(1-L_{\tt C})$ results in a periodic term with an amplitude of $L_{\tt C}A_1\simeq 24.53$~ps, which may be significant for precision timing applications. For the $P_{\tt H}$-series, there will be a periodic term $A_{\tt H1}$ similar to $A_1$, but with a slightly larger amplitude. It will contribute $L_{\tt B}A_{\tt H1} \simeq 25.69$~ps and $L_{\tt L}A_{\tt H1}  \simeq 0.05$~ps. Thus, the constants $ L_{\tt C}$ and $ L_{\tt B}$ may be kept in the model (\ref{eq:(nonN2+)2-TT}) as part of the pre-factors multiplying the periodic terms, but $L_{\tt L}$ may be neglected.  

\subsection{Explicit form of the constant and periodic terms }
\label{sec:AppB}

Eqs.~(\ref{eq:(nonN2+)2})  and (\ref{eq:(nonN2+)2-TT})  provide the relationship between {\tt TT} and {\tt TL}  using the periodic terms $P(t)$ and $P_{\tt H}(t)$. While $P(t)$ is well-known \cite{Fairhead-Bretagnon:1990,Fukushima:1995,Irwin-Fukushima:1999,Fukushima:2010}, the specific form of $P_{\tt H}(t)$ has yet to be determined. However,  as seen in  (\ref{eq:(nonN2+)2})  and (\ref{eq:(nonN2+)2-TT}), it is not the individual forms of these functions that are crucial, but rather their difference, $P(t)-P_{\tt H}(t)$.
Additionally, because the Moon orbits the Earth, there may be similar terms in the functions that could cancel out in the difference.  We will explore this possibility with the aim of simplifying the relevant expressions describing $({\tt TT} - {\tt TL})$.  

Eq.~(\ref{eq:(nonN2-in+2)}) establishes the relationship between {\tt TT} and {\tt TL} referring to  {\tt TDB} as the common time scale. Considering our target time transfer uncertainty of $10$~ps and the time rate uncertainty of $1.22 \times 10^{-16}=10.54$~ps/d, we can introduce simplifications. Furthermore, to get insight into the structure of the terms in (\ref{eq:(nonN2-in+2)}), we note  that, within the required level of accuracy, the gravitational potentials in this equation may be given only by their monopole components. 

With these realizations and using the constants and mean values of various dynamical quantities from Table \ref{tab:1}, the group of $c^{-2}$-terms present in Eq.~(\ref{eq:(nonN2-in+2)})  can be simplified and expressed as follows:
{}
\begin{eqnarray}
\delta_2 ({\tt TL} -   {\tt TT} )&=&-
\frac{1}{c^2}\int_{\tt T_0+TDB_0}^{\tt TDB}  \Big\{ \Big( {\textstyle\frac{1}{2}}v_{\tt M}^2 +  \sum_{{\tt B}\not={\tt M}}\frac{GM_{\tt B}}{r_{\tt BM}}\Big)- \Big({\textstyle\frac{1}{2}}v_{\tt E}^2 +  \sum_{{\tt B}\not={\tt E}}\frac{GM_{\tt B}}{r_{\tt BE}}\Big) \Big\} d{\tt TDB}+ \frac{1}{c^2}\Big((\vec v_{\tt E}\cdot \vec {r_{\tt EM}})-(\vec v_{\tt EM}\cdot \vec {r_{\tt M}}_{\tt TDB})\Big),~~~
\label{eq:expr2}
\end{eqnarray}
accurate to ${\cal O}\big(1.22\times 10^{-16}\big)$ in the clock rate. 

We first consider the term in (\ref{eq:expr2}) that involves the total energy at the Earth's orbit which can be written as below:
{}
\begin{eqnarray}
\frac{1}{c^2}\Big({\textstyle\frac{1}{2}}v_{\tt E}^2 +\sum_{{\tt B}\not={\tt E}}\frac{GM_{\tt B}}{r_{\tt BE}} \Big)=
\frac{1}{c^2}\Big({\textstyle\frac{1}{2}}v_{\tt E}^2 +\frac{GM_{\tt S}}{r_{\tt SE}}+\frac{GM_{\tt M}}{r_{\tt ME}}+\sum_{{\tt B}\not={\tt E,M,S}}\frac{GM_{\tt B}}{r_{\tt BE}}\Big) +{\cal O}\big(1.10\times 10^{-16}\big),
\label{eq:enE}
\end{eqnarray}
where the error term is set by the omitted $c^{-4}$-terms in the time transformations (\ref{eq:coord-tr-T1-rec}).  

To consider the Moon-related terms  in (\ref{eq:expr2}), it is instructive to express the {\tt BCRS}  position vector between a body {\tt B} and the Moon as $\vec r_{\tt BM}=\vec r_{\tt BE}+\vec r_{\tt EM}$, where $\vec r_{\tt BE}=\vec x_{\tt E}-\vec x_{\tt B}$ is the position vector from the body {\tt B} to the Earth, and  $\vec r_{\tt EM}=\vec x_{\tt M}-\vec x_{\tt E}$ is the Earth-Moon relative position vector, also $r_{\tt BM}\equiv|\vec x_{\tt BM}|$, $r_{\tt EM}\equiv|\vec x_{\tt EM}|$.  By treating $r_{\tt EM}/ r_{\tt BE}$ as a small parameter, we can express ${GM_{\tt B}}/{r_{\tt BM}} $ in the form of a series of tidal terms, as shown below, accurate to ${\cal O}({r_{\tt EM}^4}/{r_{\tt BE}^5})$:
{}
\begin{eqnarray}
 \frac{GM_{\tt B}}{r_{\tt BM}} = \frac{GM_{\tt B}}{r_{\tt BE}} - \frac{GM_{\tt B}}{r^3_{\tt BE}}(\vec r_{\tt BE}\cdot \vec r_{\tt EM}) + \frac{GM_{\tt B}}{2r^5_{\tt BE}}\Big(3(\vec r_{\tt BE}\cdot \vec r_{\tt EM})^2 -r_{\tt BE}^2r_{\tt EM}^2\Big) - \frac{GM_{\tt B}}{2r^7_{\tt BE}}(\vec r_{\tt BE}\cdot \vec r_{\tt EM})\Big(5(\vec r_{\tt BE}\cdot \vec r_{\tt EM})^2 -3r_{\tt BE}^2r_{\tt EM}^2\Big).~~~
\label{eq:expand2}
\end{eqnarray}

Using result (\ref{eq:expand2}), and representing $\vec v_{\tt M}=\vec v_{\tt E}+\vec v_{\tt EM}$, where $\vec v_{\tt E}$ is the {\tt BCRS} velocity of the Earth and $\vec v_{\tt EM}$ is the Earth-Moon relative velocity, we present the first expression under the integral sign in (\ref{eq:expr2}) as below: 
{}
\begin{eqnarray}
\frac{1}{c^2}\Big({\textstyle\frac{1}{2}}v_{\tt M}^2+  \sum_{{\tt B}\not={\tt M}}\frac{GM_{\tt B}}{r_{\tt BM}} \Big)&=&\frac{1}{c^2}\Big\{
{\textstyle\frac{1}{2}}v_{\tt E}^2+{\textstyle\frac{1}{2}}v_{\tt EM}^2 +  \frac{GM_{\tt E}-GM_{\tt M}}{r_{\tt EM}}+\sum_{{\tt B}\not={\tt E,M,S}}\frac{GM_{\tt B}}{r_{\tt BE}}+\nonumber\\
+\frac{GM_{\tt S}}{r_{\tt SE}} +\frac{GM_{\tt S}}{2r^5_{\tt SE}}\Big(3(\vec r_{\tt SE}\cdot \vec r_{\tt EM})^2 -r_{\tt SE}^2r_{\tt EM}^2\Big) &-&
 \frac{GM_{\tt B}}{2r^7_{\tt SE}}(\vec r_{\tt SE}\cdot \vec r_{\tt EM})\Big(5(\vec r_{\tt SE}\cdot \vec r_{\tt EM})^2 -3r_{\tt SE}^2 r_{\tt EM}^2 \Big)+\frac{d}{dt}(\vec v_{\tt E}\cdot \vec r_{\tt EM}) \Big\},~~~~~
\label{eq:expra=9g}
\end{eqnarray}
where we kept the solar  octupole tidal term  $\propto r_{\tt EM}^3 /r^4_{\tt SE}$. The magnitude of this term was estimated to be $\simeq 1.68\times 10^{-16}$, which is very small, but large enough to be part of the model.  Also, to the required level of accuracy, the Earth's barycentric acceleration, $\vec a_{\tt E}$, can be represented by its Newtonian part, yielding 
{}
\begin{eqnarray}
 -\sum_{{\tt B}\not={\tt E,M}}\frac{GM_{\tt B}}{r^3_{\tt BE}}(\vec r_{\tt BE}\cdot \vec r_{\tt EM})=(\vec a_{\tt E}\cdot \vec r_{\tt EM})-\frac{GM_{\tt M}}{r_{\tt EM}}, \qquad  {\rm where}
\qquad  
 \vec a_{\tt E}=-\sum_{{\tt B}\not={\tt E}} \frac{GM_{\tt B}}{r^3_{\tt BE}}\vec r_{\tt BE}.
\label{eq:accE}
\end{eqnarray}

Results (\ref{eq:enE}) and (\ref{eq:expra=9g}) provide  us with more insight on the structure of  $L_{\tt C}, P(t)$ and  $L_{\tt H}, P_{\tt H}(t)$, introduced by (\ref{eq:coord-tr-QQ}) and  (\ref{eq:coord-tr-QQM}), correspondingly. Indeed, from (\ref{eq:enE}), we have the constant rate $L_{\tt C}$ and the  periodic terms $P(t)$ given as 
{}
\begin{eqnarray}
L_{\tt C}&=&\frac{1}{c^2}\Big\{\left<{\textstyle\frac{1}{2}}v_{\tt E}^2\right> +\left<\frac{GM_{\tt S}}{r_{\tt SE}}\right>+\left<\frac{GM_{\tt M}}{r_{\tt ME}}\right>+\sum_{{\tt B}\not={\tt E,M,S}}\left<\frac{GM_{\tt B}}{r_{\tt BE}}\right>\Big\}=
1.480\,826\,85
\times 10^{-8} \approx 1.2794~{\rm ms/d},~~~~~~~~~
\label{eq:expLC+}\\
\dot P(t)&=&
\frac{1}{c^2}\Big\{{\textstyle\frac{1}{2}}v_{\tt E}^2 +\frac{GM_{\tt M}}{r_{\tt ME}}+\frac{GM_{\tt S}}{r_{\tt SE}}+\sum_{{\tt B}\not={\tt E,M,S}}\frac{GM_{\tt B}}{r_{\tt BE}}\Big\}-L_{\tt C}.
\label{eq:expPt+}
\end{eqnarray}
Similarly, we establish the functional form of the constant rate $L_{\tt H}$ and the periodic terms $P_{\tt H}(t)$ as shown below:
{}
\begin{eqnarray}
L_{\tt H}&=&\frac{1}{c^2}\Big\{\left<{\textstyle\frac{1}{2}}v_{\tt E}^2\right> +\left<{\textstyle\frac{1}{2}}v_{\tt EM}^2\right> +\left<\frac{GM_{\tt S}}{r_{\tt SE}}\right>+
{\textstyle\frac{1}{4}}GM_{\tt S}\left<\frac{r_{\tt EM}^2}{r^3_{\tt SE}}\right>+\big(GM_{\tt E}-GM_{\tt M}\big)\left<\frac{1}{r_{\tt ME}}\right>+\sum_{{\tt B}\not={\tt E,M,S}}\left<\frac{GM_{\tt B}}{r_{\tt BE}}\right>\Big\}=\nonumber\\
&=&
1.482\,536\,24
\times 10^{-8} \approx 1.2809~{\rm ms/d},~~~~~~~~~
\label{eq:expLH+}\\
\dot P_{\tt H}(t)&=&
\frac{1}{c^2}\Big\{
{\textstyle\frac{1}{2}}v_{\tt E}^2+{\textstyle\frac{1}{2}}v_{\tt EM}^2 +  \frac{GM_{\tt E}-GM_{\tt M}}{r_{\tt EM}}+\sum_{{\tt B}\not={\tt E,M,S}}\frac{GM_{\tt B}}{r_{\tt BE}}+\frac{GM_{\tt S}}{r_{\tt SE}} +\frac{GM_{\tt S}}{2r^5_{\tt SE}}\Big(3(\vec r_{\tt SE}\cdot \vec r_{\tt EM})^2 -r_{\tt SE}^2r_{\tt EM}^2\Big)-\nonumber\\
&-& 
 \frac{GM_{\tt B}}{2r^7_{\tt SE}}(\vec r_{\tt SE}\cdot \vec r_{\tt EM})\Big(5(\vec r_{\tt SE}\cdot \vec r_{\tt EM})^2 -3r_{\tt SE}^2 r_{\tt EM}^2 \Big)+\frac{d}{dt}(\vec v_{\tt E}\cdot \vec r_{\tt EM}) \Big\}-L_{\tt H}.
\label{eq:expPH+}
\end{eqnarray}
Note that to evaluate the magnitude of $L_{\tt H}$ above, we used the values from Table~\ref{tab:1}. In addition, we  estimated the mean value of the relative Earth-Moon velocity as $v_{\tt EM}= 2\pi r_{\tt EM}/(27.321\,661~{\rm d} \times 86400~{\rm s/d})= r_{\tt EM}\omega_{\tt M}=1023.15$~m/s.

Eqs.~(\ref{eq:expLC+})--(\ref{eq:expPt+}) and (\ref{eq:expLH+})--(\ref{eq:expPH+}) explicitly show the complex structure of all the quantities involved, including both constant and periodic components. Ultimately, it is necessary to evaluate these expressions and monitor their behavior for timing purposes. However, when comparing clocks in cis-lunar space to their analogues near the Earth, only the differences between these quantities are significant.  As many common terms will cancel out, the differences will have a much simpler structure.  Indeed, expressions (\ref{eq:enE}) and (\ref{eq:expra=9g}) yield the following result
 {}
\begin{eqnarray}
\frac{1}{c^2}\Big({\textstyle\frac{1}{2}}v_{\tt M}^2 +\sum_{{\tt B}\not={\tt M}}\frac{GM_{\tt B}}{r_{\tt BM}} \Big)
&-&\frac{1}{c^2}\Big({\textstyle\frac{1}{2}}v_{\tt E}^2 +\sum_{{\tt B}\not={\tt E}}\frac{GM_{\tt B}}{r_{\tt BE}}  \Big)=\frac{1}{c^2}\Big\{
{\textstyle\frac{1}{2}}v_{\tt EM}^2 +  \frac{GM_{\tt E}-2GM_{\tt M}}{r_{\tt EM}} +\frac{GM_{\tt S}}{2r^5_{\tt SE}}\Big(3(\vec r_{\tt SE}\cdot \vec r_{\tt EM})^2 -r_{\tt SE}^2r_{\tt EM}^2\Big)-\nonumber\\
 &-&
 \frac{GM_{\tt B}}{2r^7_{\tt SE}}(\vec r_{\tt SE}\cdot \vec r_{\tt EM})\Big(5(\vec r_{\tt SE}\cdot \vec r_{\tt EM})^2 -3r_{\tt SE}^2 r_{\tt EM}^2 \Big)+\frac{d}{dt}(\vec v_{\tt E}\cdot \vec r_{\tt EM}) \Big\}+{\cal O}\big(1.22\times 10^{-16}\big).~~~~~
\label{eq:expra=9}
\end{eqnarray}

Following the approach demonstrated in (\ref{eq:coord-tr-QQ}) and (\ref{eq:coord-tr-QQM}), we can present result (\ref{eq:expra=9}) in a similar functional form
{}
\begin{eqnarray}
\frac{1}{c^2}\Big({\textstyle\frac{1}{2}}v_{\tt M}^2 +\sum_{{\tt B}\not={\tt M}}\frac{GM_{\tt B}}{r_{\tt BM}} \Big)
&-&\frac{1}{c^2}\Big({\textstyle\frac{1}{2}}v_{\tt E}^2 +\sum_{{\tt B}\not={\tt E}}\frac{GM_{\tt B}}{r_{\tt BE}}  \Big)=
L_{\tt EM}+\dot P_{\tt EM}(t)+\frac{d}{dt}(\vec v_{\tt E}\cdot \vec r_{\tt EM})+{\cal O}\big(1.22\times 10^{-16}\big),
\label{eq:expra=10+}
\end{eqnarray}
where the constant rate $L_{\tt EM}\simeq L_{\tt H}-L_{\tt C}$ and the  periodic terms $\dot P_{\tt EM}(t) \simeq \dot P_{\tt H}(t)-\dot P(t)$ are given as below:
{}
\begin{eqnarray}
L_{\tt EM}&=&
\frac{1}{c^2} \Big\{\left<{\textstyle\frac{1}{2}}v_{\tt EM}^2\right> +  \big(GM_{\tt E}-2GM_{\tt M}\big)\left<\frac{1}{r_{\tt EM}}\right> +
{\textstyle\frac{1}{4}}GM_{\tt S}\left<\frac{r_{\tt EM}^2}{r^3_{\tt SE}}\right>\Big\}=
1.709\,385\times 10^{-11}=1.4769~\mu{\rm s/d},~~~~~~~~~
\label{eq:expRR1+}\\
\dot P_{\tt EM}(t)&=&
\frac{1}{c^2} \Big\{{\textstyle\frac{1}{2}}v_{\tt EM}^2-\left<{\textstyle\frac{1}{2}}v_{\tt EM}^2\right> +  \big(GM_{\tt E}-2GM_{\tt M}\big)\Big(\frac{1}{r_{\tt EM}}-\left<\frac{1}{r_{\tt EM}}\right>\Big)+
{\textstyle\frac{1}{4}}GM_{\tt S}\Big(\frac{r_{\tt EM}^2}{r^3_{\tt SE}}-\left<\frac{r_{\tt EM}^2}{r^3_{\tt SE}}\right>\Big) +\nonumber\\
&+&\frac{3GM_{\tt S}}{4r^5_{\tt SE}}\Big(2(\vec r_{\tt SE}\cdot \vec r_{\tt EM})^2 -r_{\tt SE}^2r_{\tt EM}^2\Big) -
 \frac{GM_{\tt S}}{2r^7_{\tt SE}}(\vec r_{\tt SE}\cdot \vec r_{\tt EM})\Big(5(\vec r_{\tt SE}\cdot \vec r_{\tt EM})^2 -3r_{\tt SE}^2r_{\tt EM}^2\Big)\Big\}.
\label{eq:expRR2+}
\end{eqnarray}

Result (\ref{eq:expra=10+}), together with expressions (\ref{eq:expRR1+}) and (\ref{eq:expRR2+}), provides valuable insight into the structure of the constant term \( L_{\tt EM} \) and the periodic terms \( P_{\tt EM}(t) \). These expressions can now be utilized to explicitly establish the structure of the series \( P_{\tt EM}(t) \), which may serve as a replacement for the terms \( P(t) \) and \( P_{\tt H}(t) \) in (\ref{eq:(nonN2+)2}) and (\ref{eq:(nonN2+)2-TT}). (A preliminary analysis of these quantities is presented in Appendix~\ref{sec:appC}, with results given by (\ref{eq:expRR1+-bin}) and (\ref{eq:expRR2+-bin}). In particular, \( L_{\tt EM} \)  was evaluated to be \( L_{\tt EM} =1.710\,977 \times 10^{-11}=1.4782~\mu{\rm s/d}\), as shown by (\ref{eq:expRR1+-bin}) and 
the largest term in the series \( P_{\tt EM}(t) \) is estimated to be  \( \sim0.108\,782~\mu{\rm s/d} \), contributing at the mean anomalistic  period $27.55455~{\rm d}$.  When integrated over time, this term is responsible for the oscillations of \(0.472~\mu\)s, see result  (\ref{eq:expRR2+-bin-In}).)

Finally, using   (\ref{eq:expra=10+}) in (\ref{eq:expr2}) and substituting the result in (\ref{eq:(nonN2-in+2)}), we express $({\tt TL} -   {\tt TT})$ as a function of {\tt TDB} as below
 {}
\begin{equation}
{\tt TL} -   {\tt TT} =\Big(L_{\tt G}-L_{\tt L}-L_{\tt EM}\Big) \big( {\tt TDB}-{\tt T_0}-{\tt TDB}_0\big) -L_{\tt L}({\tt T}_0-{\tt T}_{\tt LO})-\Big(P_{\tt EM}({\tt TDB})-P_{\tt EM}({\tt T_0+TDB_0} )\Big)-\frac{1}{c^2}(\vec v_{\tt EM}\cdot \vec {r_{\tt M}}_{\tt TDB}),
\label{eq:(nonN2-in+2NN)}
\end{equation}
accurate to ${\cal O}\big(1.22\times 10^{-16}({\tt TDB -T_0- TDB_0})\big)$ and where ${r_{\tt M}}_{\tt TDB}$ is the {\tt TDB}-compatible position of the lunar clock.  

In the similar manner, using  (\ref{eq:expra=10+}) in (\ref{eq:(nonN2-in+2)-TT}), we obtain the same result, but expressed as a function of {\tt TT}:
{}
\begin{eqnarray}
{\tt TL} -   {\tt TT} &=&\Big(L_{\tt G}-L_{\tt L}-L_{\tt EM}\Big)\big( {\tt TT}-{\tt T_0}\big) -L_{\tt L}({\tt T}_0-{\tt T}_{\tt LO})-\Big(P_{\tt EM}({\tt TT})-P_{\tt EM}({\tt T_0} )\Big)-\frac{1}{c^2}(\vec v_{\tt EM}\cdot \vec{\cal X}_{\tt TT}),
\label{eq:(nonN2-in+2)-TT-RR}
\end{eqnarray}
 accurate to ${\cal O}\big(1.22\times 10^{-16} ({\tt TT -T_0})\big)$ and where $\vec{\cal X}_{\tt TT}$ is the {\tt TT}-compatible lunicentric position of the lunar clock. This  result is formally identical to (\ref{eq:(nonN2-in+2NN)}), as there is no rate drift between {\tt TT} and {\tt TDB}, as shown by (\ref{eq:(nonN1T)}) and (\ref{eq:(nonN1TCB)}). Also, see Appendix~\ref{sec:appC} for preliminary evaluation of the constant and periodic terms, $L_{\tt EM}$ and $P_{\tt EM}(t)$, correspondingly.  
 
With $L_{\tt EM}=L_{\tt H}-L_{\tt C}=
1.709\,385\,5 \times 10^{-11}=1.4769~\mu$s/d,  the total constant rate between the clock on or near the lunar surface and its terrestrial analogue to the accepted level of accuracy is estimated to be  
{}
\begin{eqnarray}
L_{\tt B}-L_{\tt M}\simeq L_{\tt G}-L_{\tt L}-L_{\tt EM}=\big(60.2146-2.7121-1.4769\big)~\mu{\rm s/d}=56.0256~\mu{\rm s/d}.
\label{eq:(RR)}
\end{eqnarray}

Note that, if the smaller value for the lunar radius $R_{\tt M}$ is used in (\ref{eq:(29)}) instead of $R_{\tt MQ}$, the value of $L_{\tt L}$ is estimated to be $L_{\tt L}=3.140\,587\,7\times 10^{-11}\simeq 2.7135 \,\mu$s/d. With this value, the total rate in (\ref{eq:(RR)}) is $L_{\tt B}-L_{\tt M}=56.0242~\mu{\rm s/d}$. Also, if the selenoid value of $W_{\tt gM}=2821713.3~\mathrm{m}^2\mathrm{s}^{-2}$ from \cite{Martinec-Pec:1988} is used to determine  $L_{\tt L} = 3.139\,579\,5 \times 10^{-11} \simeq 2.7126~\mu\mathrm{s}/\mathrm{d}$, the value of $L_{\tt B}-L_{\tt M}=56.0251~\mu{\rm s/d}$. This dispersion highlights the need for further studies of the lunar constants. 
 
We observe that, practically, the clock drift with a rate of $(L_{\tt B}-L_{\tt M})/(1-L_{\tt B})=56.0256~\mu$s/day is quite significant. If unaccounted for, this rate mismatch could lead to a velocity estimate error of $ c (L_{\tt B}-L_{\tt M})/(1-L_{\tt B})=19.44$~cm/s or result in a positional error of $ c \Delta t (L_{\tt B}-L_{\tt M})/(1-L_{\tt B})\simeq 699.84~ (\Delta t /1~{\rm hr})$~m. Such potential navigational errors are significant and necessitate the establishment and maintenance of the {\tt LCRS}. This is especially important for the upcoming long-term human presence on the Moon and various anticipated exploration and industrial activities.

Expressions (\ref{eq:(nonN2-in+2NN)}) and (\ref{eq:(nonN2-in+2)-TT-RR}) may  be used to synchronize {\tt TL} and {\tt TT}. Note that these results are valid for a clock rate accuracy  of $\sim 1.22 \times 10^{-16}=10.50$~ps/d, as was shown in (\ref{eq:coord-tr-T1-recM}). If needed, they may be extended to a greater accuracy using well-established frameworks, e.g., \cite{Soffel-etal:2003} and better modeling of the relevant lunar constants and periodic terms, e.g., $L_{\tt L}$, $L_{\tt H}$, $L_{\tt M}$, $P_{\tt H}$, discussed in Sec.~\ref{sec:tm-tdb}, as well as   $L_{\tt EM}$, $P_{\tt EM}$ from (\ref{eq:expra=10+}) and given by (\ref{eq:expRR1+}) and (\ref{eq:expRR2+}), correspondingly. 

It is instructive to consider the $\dot P_{\tt EM}(t)$ term given by (\ref{eq:expRR2+}). To evaluate this term, one needs to develop a comprehensive model for the lunar orbit. It is important, to accurately identify all the forces perturbing the Moon's orbit around the Earth, including gravitational influences from the Sun, planets, and other celestial bodies. These perturbations generate variations in the Moon's osculating orbital elements, which manifest as terms at distinct frequencies \cite{Williams-Boggs:2020}. A systematic analysis of each perturbation is necessary to isolate and classify these terms, providing a detailed characterization of the frequency components in the Moon's motion. 
 
 Before that model is available, we can estimate the magnitude of  the leading term in (\ref{eq:expRR2+}). In Appendix~\ref{sec:appC}, we modeled Earth-Moon as a binary with comparable masses and obtained expression for the leading term shown by (\ref{eq:expRR2+-bin-In}). This term arises from the combined contributions of several distinct sources: the gravitational potentials of the Earth and the Moon, and the solar tidal effect on the lunar orbit with the mean anomaly of $n_{\tt M}$. Both contributions are responsible for periodic contributions at the same frequency, yielding:
 {}
\begin{eqnarray}
\dot P_{\tt EM}(t)&\simeq& \frac{1}{c^2} \Big\{ \frac{2GM_{\tt E}}{a_{\tt M}}\Big(1-\frac{M_{\tt M}}{2M_{\tt E}}\Big)(1-{\textstyle\frac{1}{8}}e^2_{\tt M})-{\textstyle\frac{1}{2}}GM_{\tt S}\frac{a_{\tt M}^2}{a^3_{\tt E}}\Big\}e_{\tt M}\cos n_{\tt M}t ~= 0.108\cos n_{\tt M}t ~~\mu{\rm s/d},
\label{eq:expRR2+-bin2-int}\\
P_{\tt EM}(t)&\simeq&
\frac{1}{c^2} \Big\{ \frac{2GM_{\tt E}}{a_{\tt M}}\Big(1-\frac{M_{\tt M}}{2M_{\tt E}}\Big)(1-{\textstyle\frac{1}{8}}e^2_{\tt M})-{\textstyle\frac{1}{2}}GM_{\tt S}\frac{a_{\tt M}^2}{a^3_{\tt E}}\Big\}\frac{e_{\tt M}}{n_{\tt M}}\sin n_{\tt M}t =
 0.472 \, \sin n_{\tt M}t ~~\mu{\rm s}.
\label{eq:P-int}
\end{eqnarray}

Alternatively, we may evaluate $P_{\tt EM}(t)$ from  (\ref{eq:P-int}) by employing series expansions for the lunar orbit, as provided in \cite{Chapront_etal_2002}, which include polynomial expressions for the orbit at various lunar angles. This approach may offer valuable insights. For the required parameters, we utilize results from JPL's analysis of LLR data, which combines numerical integration of the Moon's orbit and dynamical partial derivatives with Keplerian elements and series expansions. Table~\ref{tab:4} summarizes the Moon's Keplerian elements, mean distance, and key orbital angles and periods. The lunar orbit precesses near the ecliptic due to dominant solar perturbations, which outweigh the effects of Earth's $J_2$. 

%====================================
\begin{table}[t!]
\caption{Lunar orbit and angles from \cite{Williams-etal:2009}.} 
{\begin{tabular}{rcc} \hline
Parameter & Symbol & Value \\\hline
Mean distance 	& $\left<r_{\tt M} \right>$ & 385,000.5 km \\
Semimajor axis	& ~~$a_{\tt M} = 1/\left<1/r_{\tt M}\right>$ ~~ & 384,399.0 km \\
Eccentricity	& $e_{\tt M}$ &	0.0549006\\
Mean Longitude	& $L$ &	27.32166 d \\
Mean Anomaly	& $\ell$ &	27.55455 d \\
Mean Argument of Latitude & $F$ & 27.21222 d \\
Mean Elongation of Moon from Sun & $D$ & 29.53059 d \\
\hline
\end{tabular} \label{tab:4}}
\end{table}
%====================================

The lunar orbit experiences significant perturbations due to the Sun's gravitational influence. Accurate series representations of the lunar orbit have been developed using advanced computational techniques, as detailed in \cite{Chapront-Touze_Chapront_1988,Chapront-Touze_Chapront_1991}. These series provide the dominant terms for the radial coordinate (in kilometers), given by:
{}
\begin{equation}
r_{\tt EM} ~=~ 385001 - 20905 \cos \ell - 3699 \cos(2D-\ell) - 2956 \cos 2D - 570 \cos 2\ell + 246 \cos(2\ell-2D) + ... + 109 \cos D + ... ~
\label{eq:rexp}
\end{equation}

The constant term on the right-hand side represents the Earth-Moon mean distance, which slightly larger than the semi-major axis. The $\ell$ and $2\ell$ terms correspond to elliptical contributions, while the remaining terms originate from solar perturbations. The amplitudes of the solar perturbation terms are determined by the gravitational parameters of the Earth, Moon, and Sun, as well as the dynamics of the lunar orbit and the Earth-Moon system’s motion around the Sun. The periods of the terms in (\ref{eq:rexp}), listed in order, are 27.555 d, 31.812 d, 14.765 d, 13.777 d, 205.9 d, and 29.531 d, ensuring distinct and well-separated frequencies.

Using expression (\ref{eq:rexp}), we estimate the leading terms in the series $\dot P_{\tt EM}(t)$  from (\ref{eq:expRR2+-bin2}) and $ P_{\tt EM}(t)$  from (\ref{eq:expRR2+-bin-In}), as below  
{}
\begin{eqnarray}
\dot P_{\tt EM}(t)&\simeq&-
\frac{1}{c^2} \Big\{\frac{2GM_{\tt E}}{\left<r_{\tt M}\right>}\Big(1-\frac{M_{\tt M}}{2M_{\tt E}}\Big)-{\textstyle\frac{1}{2}}GM_{\tt S}\frac{\left<r_{\tt M}\right>^2}{a^3_{\tt E}}\Big\}\, 0.0543 \cos \ell ~\simeq
0.107\cos \ell ~\mu{\rm s/d},
\label{eq:expRR2+-bin2-LLR}\\
P_{\tt EM}(t)&=&-
\frac{1}{c^2} \Big\{\frac{2GM_{\tt E}}{\left<r_{\tt M}\right>}\Big(1-\frac{M_{\tt M}}{2M_{\tt E}}\Big)-{\textstyle\frac{1}{2}}GM_{\tt S}\frac{\left<r_{\tt M}\right>^2}{a^3_{\tt E}}\Big\}\frac{0.0543 \sin \ell}{2\pi/{\rm 27.55455~d}}
\simeq 0.470 \, \sin \ell~~\mu{\rm s}.~~~
\label{eq:expRR2+-bin-In-LLR}
\end{eqnarray}
These results closely align with (\ref{eq:expRR2+-bin2-int}) and (\ref{eq:P-int}). Establishing the behavior of these quantities at other orbital frequencies will be crucial for accurate time transfer (\ref{eq:(nonN2-in+2)-TT-RR}). The approach outlined here can be effectively applied for this purpose.

As a result, in addition to all the factors mentioned in Sec.~\ref{sec:tech-basis}, synchronization between {\tt TL} and {\tt TT} is another requirement for establishing the {\tt LCRS}, especially in the context of long-term lunar operations. As shown in (\ref{eq:(nonN2-in+2)-TT-RR})--(\ref{eq:(RR)}), even at the origin of the {\tt LCRS},  clocks on the Moon experience a relativistic drift of  $\sim56.0256~\mu$s/d relative to {\tt TT} taken at the geocenter, necessitating periodic corrections to maintain synchronization. In addition, as seen from (\ref{eq:expRR2+-bin-In-LLR}), there are also a periodic variations with the largest amplitude of 0.470~$\mu$s from (\ref{eq:expRR2+-bin-In-LLR}), occurring at the mean anomaly period of $27.554\,55~{\rm d}$. Therefore, a lunar time scale {\tt TL}, incorporating relativistic corrections and adjustments for tidal and libration effects, is essential for the {\tt LCRS} to maintain temporal alignment with terrestrial systems.

\section{Discussion}
\label{sec:conlude}

The growing demand for precise positioning, navigation, and timing (PNT) services in lunar exploration highlights the necessity of establishing a Luni-centric Coordinate Reference System ({\tt LCRS}) with Luni-centric Coordinate Time ({\tt TCL}) at its origin and Lunar Time ({\tt TL}) on the lunar surface. These systems are crucial for supporting operational efficiency, scientific endeavors, and future commercial activities on the Moon. Existing Earth-centric frameworks are inadequate for these demands, necessitating the development of an independent lunar coordinate and time system.

A standardized lunar time scale, {\tt TL}, is essential for synchronizing activities and operations involving multiple landers, rovers, and orbiters. Establishing a common time reference ensures coordination between assets, prevents conflicts, and enhances collaboration across missions. Accurate timing also underpins reliable Earth-Moon communication, ensuring precise data transmission and reception. This synchronization is particularly vital for autonomous systems that require robust timing frameworks for uninterrupted operations.

The implementation of a dedicated {\tt LCRS} is equally critical for navigation on the lunar surface. Unlike Earth, the Moon's environment presents significant challenges, including highly irregular terrain and the absence of a global magnetic field. These factors necessitate a reference system that enables precise positioning for landers, rovers, and other assets. For instance, targeting specific landing sites or navigating polar regions for resource utilization, such as water ice extraction, requires sub-meter positional accuracy. The {\tt LCRS} facilitates these capabilities while ensuring the safety and reliability of surface operations.

For scientific applications, an {\tt LCRS} provides the essential geospatial context needed for the accurate placement of instruments, thereby enhancing our understanding of the Moon's geology and environment. Precise navigation allows rovers to traverse the lunar surface efficiently, avoiding obstacles and reaching designated sites. This precision not only extends the operational life of the rovers but also maximizes the scientific return from each mission.

Defining the selenoid, the lunar equivalent of Earth’s geoid, presents substantial challenges for accurately specifying $L_{\tt L}$, given by (\ref{eq:(29)}) and  $L_{\tt M}$, given by (\ref{eq:(nonN4)}) and (\ref{eq:consTM}). These challenges stem from the Moon’s highly irregular gravitational field, which exhibits significant spatial and temporal variability due to mascons, tidal effects, and librational motions. Unlike Earth’s geoid, which benefits from the presence of a globally uniform reference surface (sea level), the lunar selenoid is affected by its uneven topography and dynamic gravitational anomalies, complicating its characterization.

These complexities directly impact the establishment of precise {\tt LCRS}. Temporal variability in the Moon’s gravity harmonics, driven by tidal and libration-induced shifts, introduces positional uncertainties that necessitate continual adjustments to coordinate definitions. For example, Earth-induced tidal effects can cause spatial displacements of up to 20~m over extended periods, requiring detailed modeling of these variations. 

From a relativistic perspective, the orbital energy difference between Earth and the Moon introduces a relativistic secular drift in the $({\tt TL-TT})$ transformation of $L_{\tt B}-L_{\tt M}\simeq L_{\tt G}-L_{\tt L}-L_{\tt EM}\sim 56.0256~\mu$s/day (\ref{eq:(RR)}) with a periodic term with the leading contribution of 0.470 $\mu$s at the mean anomalistic period (\ref{eq:expRR2+-bin-In-LLR}). Further research is needed to refine the definitions of $L_{\tt L}$ and  $L_{\tt M}$, accounting for the intricate interplay of lunar gravitational harmonics and librational dynamics. With the fidelity of gravitational potential models (e.g., GRGM1200A), introducing uncertainties on the order of $10^{-16}$ in timing calculations, this effort is critical to ensuring that the selenoid provides a stable and reliable geospatial framework for lunar navigation and exploration. As a result, this necessitates the definition of {\tt TL} that incorporates relativistic corrections and accounts for periodic variations caused by tidal and libration-induced effects. 

Time synchronization across the Earth-Moon system is critical for supporting operations requiring precise coordination, such as autonomous navigation, high-precision time transfer, and scientific measurements \cite{Bar-Sever-etal:2024}. Future lunar infrastructure, including navigation beacons and lunar bases, will rely on standardized lunar time and coordinate systems to ensure seamless operation. Accurate location tracking and time synchronization are fundamental for maintaining operational safety and enabling scalable lunar operations.

Implementing {\tt TL} and {\tt LCRS} on the lunar surface poses challenges, including ensuring clock stability under extreme temperature fluctuations, maintaining real-time synchronization across Earth-Moon distances, and integrating these systems with terrestrial and cislunar infrastructures. Nevertheless, these systems are foundational for future lunar infrastructure, such as navigation beacons, bases, and commercial operations. Standardized time and coordinate systems will enable safe and efficient transportation, resource management, and international collaboration.

In conclusion, while this paper outlines the foundational framework and equations for a fully relativistic {\tt LCRS}, further work is required to refine the specific relativistic time and position transformations for diverse users across cislunar space, including landers, orbiters, and missions at Earth-Moon Lagrange points. These refinements will ensure that the {\tt LCRS} supports a broad range of applications, including PNT services, astronomy, lunar geology, and fundamental physics. This work is ongoing, and the results will be presented in future studies.

\section*{Acknowledgments}

The work described here was carried out the Jet Propulsion Laboratory, California Institute of Technology, Pasadena, California, under a contract with the National Aeronautics and Space Administration.
% \textcopyright 2024. California Institute of Technology. Government sponsorship acknowledged.
 
%\bibliography{moon-time}

\appendix

\section{Planet-Caused Relativity }
\label{sec:planet-GR}

The gravity of distant planets at the Moon causes a relativistic effect. This effect is accounted for in the definitions of the constants $L_{\tt B}$ and $L_{\tt M}$, given by the expressions (\ref{eq:(nonN3)}) and (\ref{eq:(nonN4)}), correspondingly. We ignore the displacement of the Moon from the Earth-Moon barycenter, 1\% of the distance to Venus when it is closest, and calculate the positive potential at the Earth-Moon barycenter:
{}
\begin{align}
U_{\tt BP} = \frac{GM_{\tt P}}{|\vec r_{\tt E}-\vec r_{\tt P}|},
\label{eq:(nonN5)}
\end{align}
where $M_{\tt P}$ is the planet’s mass and $|\vec r_{\tt E}-\vec r_{\tt P}|$ is its distance from the Earth-Moon barycenter. The relativistic effect is then $U_{\tt BP}/c^2$. 
{}
\begin{align}
|\vec r_{\tt E}-\vec r_{\tt P}|^2 = r_{\tt E}^2 + r_{\tt P}^2  -  2 r_{\tt E} r_{\tt P} \cos S,
\qquad {\rm where} \qquad
\cos S = \cos u_{\tt E} \cos u_{\tt P} + \cos i_{\tt P} \sin u_{\tt E} \sin u_{\tt P},
\label{eq:(nonN7)}
\end{align}
and $u_{\tt E}$ and $u_{\tt P}$ are the arguments of latitude measure from the node of the planet’s orbit plane on the ecliptic plane and $i_{\tt P}$ is the orbit plane’s inclination. 
{}
\begin{align}
u_{\tt E} = \varpi_{\tt E}  -  \Omega_{\tt P} + f_{\tt E}
\qquad {\rm and} \qquad
u_{\tt P} = \omega_{\tt P} + f_{\tt P},
\label{eq:(nonN9)}
\end{align}
where $\varpi_{\tt E}$  is the Earth-Moon barycenter longitude of perihelion, $\omega_{\tt P}$ is the planet’s argument of latitude, and $\Omega_{\tt P}$ is the planet’s orbit plane node on the ecliptic plane. The two true anomalies are $f_{\tt E}$ and $f_{\tt P}$ for the Earth-Moon barycenter and planet, respectively. The radii are computed from 
{}
\begin{align}
r = \frac{a (1 - e^2)}{ 1 + e \cos f }, 
\label{eq:(nonN10)}
\end{align}
with the two subscripts. The semi-major axis is $a$ and the eccentricity is $e$. 

\begin{table*}[t!]
\vskip-15pt
\caption{Relativistic effects on the rates at Earth-Moon barycenter from the planets at J2000.
\label{tab:2}}
\begin{tabular}{|c| c | c|}\hline
Planet 	& $e_{\tt E}=e_{\tt P}=i_{\tt P}=0$ & Fourier \\
 	& $U_{\tt BP}/c^2$, $\times10^{-15}$ & 
$U_{\tt BP}/c^2$, $\times10^{-15}$
  \\\hline
Mercury 	& ~~~~~1.706	& ~~~~1.710 \\
Venus	& ~~~28.829	& ~~~28.772 \\
Mars	        & ~~~~2.395	& ~~~~2.401 \\
Jupiter	& 1828.565	& 1828.622 \\
Saturn	& ~~296.146	 & ~~296.148 \\
Uranus	& ~~~22.440	& ~~~22.440 \\
Neptune	& ~~~16.892	& ~~~16.892 \\
Pluto	         & ~~~~0.002	& ~~~~0.002 \\\hline
Total 	& 2196.975      & 2196.986 \\
\hline
\end{tabular}
\end{table*}

The relativistic effects from the planets’ potentials was calculated in two ways. The first way is an approximation that assumes two circular ($e_{\tt E} = e_{\tt P} = 0$) coplanar ($i_{\tt P} = 0$) orbits. Then the time averaged potential $\left<U_{\tt P}\right>$ at the Earth depends on a complete elliptic integral of the first kind $K(k)$ \cite{Abramovitz-Stegun:1965}, where $k$ is the modulus. Then 
{}
\begin{align}
\left<U_{\tt P}\right> &= \frac{GM_{\tt P}}{(a_{\tt E} + a_{\tt P})} \frac{2}{\pi} K(k),
\label{eq:(nonN11a)}
\qquad 
k = \frac{2 \sqrt{a_{\tt E} a_{\tt P}}}{( a_{\tt E} + a_{\tt P} )},
\end{align}
where $a_{\tt E}$ and $a_{\tt P}$ are the semi major axes of the Earth and planet, respectively. 

The second technique is more exact. Elliptical orbits with a mutual inclination $(i_{\tt P})$ are used. The potential Eq.~(\ref{eq:(nonN11a)}) is Fourier analyzed. Integrals over time, or equivalently mean anomaly, can be transformed into integrals over true anomaly by using 
{}
\begin{align}
dM = \frac{(r/a)^2}{(1 - e^2)^\frac{1}{2}}df,
\label{eq:(nonN12)}
\end{align}
with the two subscripts. Then the constant part of a planet’s potential at the Earth-Moon barycenter is 
{}
\begin{align}
\left<U_{\tt P}\right> = GM_{\tt P} \frac{1}{(2\pi)^2} \int_0^{2\pi}\int_0^{2\pi} \frac{(r_{\tt E}/a_{\tt E})^2 (r_{\tt P}/a_{\tt P})^2}{\Delta(1 - e_{\tt E}^2)^\frac{1}{2} (1 - e_{\tt P}^2)^\frac{1}{2}} df_{\tt E} df_{\tt P}.
\label{eq:(nonN13)}
\end{align}

The planetary $GM_{\tt P}$ are from \cite{Park-etal:2021} and the mean elements at J2000 are from \cite{Simon-etal:1994}. Both techniques give similar results with differences of 0.2\% for Mercury and Venus and 0.4\% for Mars. The Fourier-derived constants are in Table~\ref{tab:2}. The two totals for all planets differ by 0.0005\%. We note that the planetary elements are changing with time, so there will be small changes away from J2000. The values in Table~\ref{tab:2} may be compared with those in Table 2 of  \cite{Fukushima:2010}.

\section{Relevant Resolutions of the IAU}
\label{sec:appA}

There have been a series of IAU resolutions that relate to time and space transformations \cite{Soffel-etal:2003,Petit-Luzum:2010}. Below, we will summarize some of the most relevant ones (see also \url{http://www.iaufs.org/res.html}):
\vspace{4pt}

{\bf IAU 2000 Resolution B1.3:} Defines the Barycentric Celestial Reference System ({\tt BCRS}) and Geocentric Celestial Reference System ({\tt GCRS}). Harmonic coordinates are used for both. Gives the time transformation. 
\vspace{4pt}

{\bf IAU 2000 Resolution B1.5:} 
Discusses ${\tt TCB}  -  {\tt {\tt TCG}}$. Specifies $\left <d{\tt {\tt TCG}}/d{\tt TCB}\right> = 1  -  L_{\tt C}$ and $\left<d{\tt TT}/d{\tt TCB}\right> = 1  -  L_{\tt B}$. Gives 
$L_{\tt C} = 1.48082686741\times 10^{-8}\pm 2\times 10^{-17}$ and $L_{\tt B} = 1.55051976772\times 10^{-8} \pm 2\times 10^{-17}$. Specifies relation with $L_{\tt G}$, namely $1  -  L_{\tt B} = ( 1  -  L_{\tt C} ) ( 1  -  L_{\tt G} ).$ Consequently, $L_{\tt B} = L_{\tt C} + L_{\tt G}  -  L_{\tt C} L_{\tt G}$, 
$L_{\tt C} = ( L_{\tt B}  -  L_{\tt G} )/( 1  -  L_{\tt G} )$, $L_{\tt G} = ( L_{\tt B}  -  L_{\tt C} )/( 1  -  L_{\tt C} )$, $1  -  L_{\tt G} = ( 1  -  L_{\tt B} )/( 1  -  L_{\tt C} )$. Says 
$({\textstyle\frac{1}{2}} v^2_{\tt E} + \sum_{\tt B\not=E}U_{\tt B} )_{{\tt TCB}} = 1/( 1  -  L_{\tt B} ) ( {\textstyle\frac{1}{2}}v^2_{\tt E} + \sum_{\tt B\not=E}U_{\tt B} )_{\tt TDB}$. 
\vspace{4pt}

{\bf IAU 2000 Resolution B1.9:} 
Redefines {\tt TT}. Specifies $d{\tt TT}/d{\tt {\tt TCG}} = 1  -  L_{\tt G}$. Value of $L_{\tt G} = 6.969290134\times 10^{-10}$ is given as a defining constant. 
\vspace{4pt}

{\bf IAU 2006 Resolution B3:} 
Gives ${\tt TDB} = {\tt TCB}  -  L_{\tt B} ( {\tt TCB}  -  {\tt T}_0 ) 86400 + {\tt TDB}_0$, ${\tt T}_0$ is ${\rm MJD} 43144 + 32.184~{\rm s}$, 
$L_{\tt B} = 1.550519768\times 10^{-8}$, and ${\tt TDB}_0 = -65.5~\mu$s. 
\vspace{4pt}

{\bf IAU 2009 Resolution B2:} 
Current best estimates of astronomical constants are may be found at the following website: \url{https://maia.usno.navy.mil/NSFA/CBE.html }

\section{Approximate structure of $L_{\tt EM}$ and $\dot P_{\tt EM}$}
\label{sec:appC}

In Section \ref{sec:AppB}, we analyzed the term (\ref{eq:expra=9}), which quantifies the difference between the orbital energies of the Moon and the Earth in their motion around the Sun. That expression resulted in (\ref{eq:expra=10+}) which has the same meaning for the lunar orbit around the Earth as (\ref{eq:coord-tr-QQ}) and (\ref{eq:coord-tr-QQM}) for the Earth and the Moon around the Sun, correspondingly.  

The constant \( L_{\tt EM} \) from (\ref{eq:expRR1+}) and the series of small periodic terms \( \dot{P}_{\text{EM}} \) from (\ref{eq:expRR2+}) define the time transformations between the {\tt GCRS} and {\tt LCRS}. A comprehensive modeling and analysis effort is required to derive the analytical structure and to evaluate the magnitudes and long-term behavior of each term, analogous to the methodology used for the constant \( L_{\tt C} \) and the time series \( P(t) \) in (\ref{eq:coord-tr-QQ}) developed for Earth's motion relative to the {\tt BCRS} (e.g., \cite{Fairhead-Bretagnon:1990, Fukushima:1995, Irwin-Fukushima:1999, Fukushima:2010}).

Before the results of such an analysis become available, one can estimate the magnitudes of the leading terms in these quantities. To do so, we model the Earth-Moon system as a binary, with the lunar orbit expressed as \( r_{\tt EM} = a_{\tt M}(1 - e_{\tt M}\cos E_{\tt M}) \), where \( a_{\tt M} \) is the lunar semi-major axis,  $e_{\tt M} = 0.0549006$ is the orbital eccentricity, and \( E_{\tt M} \) is the eccentric anomaly. The temporal dependence of \( r_{\tt EM} (t)\) follows from the standard relation \( E_{\tt M}(t) - e_{\tt M} \sin E_{\tt M}(t) = n_{\tt M}\big(t - t_0\big) \), where \( {\cal M}_{\tt M} = n_{\tt M}\big(t - t_0\big) \) is the lunar mean anomaly, which may also include a constant. The mean motion, \( n_{\tt M} \), is given by  $n_{\tt M}=\sqrt{G(M_{\tt E}+M_{\tt M})/a^3_{\tt M}}$, where \( M_{\tt E} \) and \( M_{\tt M} \) are the masses of the Earth and Moon, respectively (e.g., \cite{Montenbruck-Gill:2012,Poisson-Will:2014}). With such parameterization, the relative Moon-Earth orbital velocity term is given by the equation:
{}
\begin{eqnarray}
{\textstyle\frac{1}{2}}v_{\tt EM}^2= {\textstyle\frac{1}{2}}G(M_{\tt E}+M_{\tt M})\Big(\frac{2}{r_{\tt EM}}-\frac{1}{a_{\tt M}}\Big).
\label{eq:vel-EM}
\end{eqnarray}
Substituting this in (\ref{eq:expra=9}), we present the resulting expression as below:
 {}
\begin{eqnarray}
\frac{1}{c^2}\Big({\textstyle\frac{1}{2}}v_{\tt M}^2 +\sum_{{\tt B}\not={\tt M}}\frac{GM_{\tt B}}{r_{\tt BM}} \Big)
&-&\frac{1}{c^2}\Big({\textstyle\frac{1}{2}}v_{\tt E}^2 +\sum_{{\tt B}\not={\tt E}}\frac{GM_{\tt B}}{r_{\tt BE}}  \Big)=\frac{1}{c^2}\Big\{
-\frac{G(M_{\tt E}+M_{\tt M})}{2a_{\tt M}}+  \frac{2GM_{\tt E}-GM_{\tt M}}{r_{\tt EM}} +
{\textstyle\frac{1}{4}}GM_{\tt S}\frac{r_{\tt EM}^2}{r^3_{\tt SE}}+\nonumber\\
+\, \frac{3GM_{\tt S}}{4r^5_{\tt SE}}\Big(2(\vec r_{\tt SE}\cdot \vec r_{\tt EM})^2 &-& r_{\tt SE}^2r_{\tt EM}^2\Big)-
 \frac{GM_{\tt B}}{2r^7_{\tt SE}}(\vec r_{\tt SE}\cdot \vec r_{\tt EM})\Big(5(\vec r_{\tt SE}\cdot \vec r_{\tt EM})^2 -3r_{\tt SE}^2 r_{\tt EM}^2 \Big)+\frac{d}{dt}(\vec v_{\tt E}\cdot \vec r_{\tt EM}) \Big\}+{\cal O}\big(1.22\times 10^{-16}\big).~~~~~
\label{eq:expra=9-bin}
\end{eqnarray}

We model Earth's orbit around the Sun as \( r_{\tt SE} = a_{\tt E}(1 - e_{\tt E}\cos E_{\tt E}) \), where \( a_{\tt E} \) is the semi-major axis,  $e_{\tt E} = 0.0167086$ is its orbital eccentricity, and \( E_{\tt E} \) is the eccentric anomaly. Also, we define the angle \( \phi_{\tt SM} = \angle (\vec{n}_{\tt SE}, \vec{n}_{\tt EM}) \), such that \( (\vec{r}_{\tt SE} \cdot \vec{r}_{\tt EM}) \simeq r_{\tt SE} r_{\tt EM} \cos (E_{\tt E}-E_{\tt M}) \). Considering  the long-term averages of the inverse distances \( r_{\tt EM} \) and \(r_{\tt SE}\) are given by \( \big<1/r_{\tt EM}\big> = 1/a_{\tt M} \) and  \( \big<1/r_{\tt SE}\big> = 1/a_{\tt E} \), we average (\ref{eq:expra=9-bin}) to derive expression for \( L_{\tt EM} \), as introduced in (\ref{eq:expRR1+}):
{}
\begin{eqnarray}
L_{\tt EM}&=&\frac{1}{c^2}\Big\{
-\frac{G(M_{\tt E}+M_{\tt M})}{2a_{\tt M}}+ (2GM_{\tt E}-GM_{\tt M}) \Big<\frac{1}{r_{\tt EM}}\Big> +
{\textstyle\frac{1}{4}}GM_{\tt S}\Big<\frac{r_{\tt EM}^2}{r^3_{\tt SE}}\Big>\Big\}=\nonumber\\
&=&
\frac{1}{c^2} \Big\{{\textstyle\frac{3}{2}} \frac{GM_{\tt E}}{a_{\tt M}}\Big(1-\frac{M_{\tt M}}{M_{\tt E}}\Big) +
{\textstyle\frac{1}{4}}GM_{\tt S}\frac{a_{\tt M}^2}{a^3_{\tt E}}\Big\}=
1.710\,977 \times 10^{-11}=1.4782~\mu{\rm s/d}.~~~~~~~~~
\label{eq:expRR1+-bin}
\end{eqnarray} 
Note that the value of \( L_{\tt EM} \) in (\ref{eq:expRR1+-bin}) is slightly higher than that in (\ref{eq:expRR1+}), which was based on the mean lunar velocity of 1023.15~m/s, given after (\ref{eq:expPH+}). Using the value of \( L_{\tt EM} \) from (\ref{eq:expRR1+-bin}), the total rate in (\ref{eq:(RR)}) becomes \( L_{\tt B} - L_{\tt M} = 56.0243~\mu{\rm s/d} \). Although the difference in the estimated values of \( (L_{\tt B} - L_{\tt M}) \) is small, it warrants a dedicated study.

With (\ref{eq:expRR1+-bin}) in hand, we  now develop expression for  \( \dot{P}_{\text{EM}} \) given by (\ref{eq:expRR2+}),  which takes the form:
\begin{eqnarray}
\dot P_{\tt EM}(t)&=&
\frac{1}{c^2} \Big\{ \big(2GM_{\tt E}-GM_{\tt M}\big)\Big(\frac{1}{r_{\tt EM}}-\left<\frac{1}{r_{\tt EM}}\right>\Big)+
{\textstyle\frac{1}{4}}GM_{\tt S}\Big(\frac{r_{\tt EM}^2}{r^3_{\tt SE}}-\left<\frac{r_{\tt EM}^2}{r^3_{\tt SE}}\right>\Big) +\frac{3GM_{\tt S}}{4r^5_{\tt SE}}\Big(2(\vec r_{\tt SE}\cdot \vec r_{\tt EM})^2 -r_{\tt SE}^2r_{\tt EM}^2\Big) \Big\}=\nonumber\\
 &=&
\frac{1}{c^2} \Big\{\frac{2GM_{\tt E}}{a_{\tt M}}\Big(1-\frac{M_{\tt M}}{2M_{\tt E}}\Big)\frac{e_{\tt M}\cos E_{\tt M}}{1-e_{\tt M}\cos E_{\tt M}}+
{\textstyle\frac{1}{4}}GM_{\tt S}\frac{a_{\tt M}^2}{a^3_{\tt E}}\Big(3e_{\tt E}\cos E_{\tt E} -2e_{\tt M}\cos E_{\tt M}
+3\cos2 (E_{\tt M} -E_{\tt E} )\Big)\Big\},
\label{eq:expRR2+-bin}
\end{eqnarray} 
where (\ref{eq:expRR2+-bin}) is accurate to ${\cal O}\big(1.68\times 10^{-16}\big)$, which is set by the omitted 
term $\propto {\cal O}(r_{\tt EM}^3/r^4_{\tt SE})$ in (\ref{eq:expRR2+}). 

To integrate (\ref{eq:expRR2+-bin}), we first expand it as a power series in eccentricity, while keeping the most significant terms
{}
\begin{eqnarray}
\dot P_{\tt EM}(t)&\simeq&
\frac{1}{c^2} \Big\{\frac{2GM_{\tt E}}{a_{\tt M}}\Big(1-\frac{M_{\tt M}}{2M_{\tt E}}\Big)\Big(e_{\tt M}(1-{\textstyle\frac{1}{8}}e^2_{\tt M})\cos n_{\tt M}t +e^2_{\tt M}\cos 2n_{\tt M}t+{\textstyle\frac{9}{8}}e^3_{\tt M}\cos 3n_{\tt M}t\Big)+
\nonumber \\
&+&
{\textstyle\frac{1}{4}}GM_{\tt S}\frac{a_{\tt M}^2}{a^3_{\tt E}}\Big(3e_{\tt E}\cos n_{\tt E}t-2e_{\tt M}\cos n_{\tt M}t +3\cos2 (n_{\tt M}- n_{\tt E})t\Big) +{\cal O}(2.08\times 10^{-16})\Big\}\simeq\nonumber\\
 &\simeq&
\Big(0.1086\cos n_{\tt M}t +
0.0060\cos 2n_{\tt M}t+
0.0042\cos2 (n_{\tt M}-n_{\tt E})t+{\cal O}(4\times 10^{-4})\Big)~~\mu{\rm s/d},
\label{eq:expRR2+-bin2}
\end{eqnarray}
where the error comes from the $\propto {\cal O}(e^3_{\tt M}\cos 3n_{\tt M}t)$ contribution of the first term.

Integrating result (\ref{eq:expRR2+-bin2}) over time, provides the leading periodic terms in the series $P_{\tt EM}(t)$ given below
{}
\begin{eqnarray}
P_{\tt EM}(t)&=&\int \dot P_{\tt EM}(t) dt\simeq
\frac{1}{c^2} \Big\{\frac{2GM_{\tt E}}{a_{\tt M}n_{\tt M}}\Big(1-\frac{M_{\tt M}}{2M_{\tt E}}\Big)\Big(e_{\tt M}(1-{\textstyle\frac{1}{8}}e^2_{\tt M})\sin n_{\tt M}t+{\textstyle\frac{1}{2}}e^2_{\tt M}\sin 2n_{\tt M}t\Big)
+\nonumber\\
 &+&
 {\textstyle\frac{1}{4}}GM_{\tt S}\frac{a_{\tt M}^2}{a^3_{\tt E}}\Big(\frac{3e_{\tt E} }{n_{\tt E}}\sin n_{\tt E}t-\frac{2e_{\tt M}}{n_{\tt M}} \sin n_{\tt M}t+
\frac{3}{2(n_{\tt M}- n_{\tt E})}\sin2 (n_{\tt M} - n_{\tt E})t\Big)\Big\}+
{\cal O}(0.5~{\rm ns})\simeq\nonumber\\
&\simeq&\Big(
0.472 \, \sin n_{\tt M}t+0.013\, \sin 2n_{\tt M}t+
0.004\,\sin n_{\tt E}t+0.009\, \sin2 (n_{\tt M}- n_{\tt E})t+{\cal O}(5\times 10^{-4})\Big)~~\mu{\rm s}.~~~
\label{eq:expRR2+-bin-In}
\end{eqnarray}

An accurate solution to $P_{\tt EM}(t)$ would consist of a series of small terms with varying magnitudes and periodicities. Properly identifying all contributions requires accounting for the diverse forces perturbing the Moon's orbit, including gravitational influences from the Sun, Earth, and other planets. A systematic analysis of these perturbations is essential to isolate and classify the terms, enabling a comprehensive characterization of all significant contributions.

\end{document}